\documentclass[a4paper,12pt,authoryear]{article}
\usepackage{amsmath,amsfonts,amssymb,epsfig}
\usepackage{fullpage}
\usepackage{natbib}

\usepackage{color}

\newcommand{\be}{\begin{equation}}
\newcommand{\en}{\end{equation}}

\def \bm#1{\mbox{\boldmath{$#1$}}}   % this is used to write boldface Greek

\def \tr{\mbox{tr\hskip 1pt}}
\def \div{\mbox{div\hskip 1pt}}
\def \Div{\mbox{Div\hskip 1pt}}
\def \grad{\mbox{grad\hskip 1pt}}
\def \Grad{\mbox{Grad\hskip 1pt}}

\begin{document}

\numberwithin{equation}{section}

%+++++++++++++++++++++++++++++++++++++

\title{Initial stresses in elastic solids: \\
Constitutive laws and acoustoelasticity}
%++++++++++++++++++++++++++++++++++++++

\author{Moniba Shams$^a$, Michel Destrade$^{b}$, Ray W. Ogden$^{a}$,\\[12pt]
 $^a$School of Mathematics and Statistics,  \\
 University of Glasgow, \\ University Gardens, Glasgow G12 8QW,  Scotland, UK; \\[12pt]
 $^b$School of Mathematics, Statistics and Applied Mathematics,  \\
   National University of Ireland Galway, \\ University Road, Galway, Ireland.}

\date{}

%++++++++++++++++++++++++++++++++++++++++++++++++++++++++++
\maketitle
%++++++++++++++++++++++++++++++++++++++++++++++++++++++++++

%%%%%%%%%%%%%%%%%%

\begin{abstract}

On the basis of the nonlinear theory of elasticity, the general constitutive equation for an isotropic hyperelastic solid in the presence of initial stress is derived.
This derivation involves invariants that couple the deformation with the initial stress and in general, for a compressible material, it requires 10 invariants, reducing to 9 for an incompressible material.   Expressions for the Cauchy and nominal stress tensors in a finitely deformed configuration are given along with the elasticity tensor and its specialization to the initially stressed undeformed configuration.  The equations governing infinitesimal motions superimposed on a finite deformation are then used to study the combined effects of initial stress and finite deformation on the propagation of  homogeneous plane waves in a homogeneously deformed and initially stressed solid of infinite extent.  This general framework allows for various different specializations, which make contact with earlier works.  In particular, {\color{black} connections with} results derived within Biot's classical theory are highlighted.  The general results are also specialized to the case of a small initial stress and a small pre-deformation, i.e. to the evaluation of the acoustoelastic effect.  Here the formulas derived for the wave speeds cover the case of a second-order elastic solid without initial stress and subject to a uniaxial tension [Hughes and Kelly, Phys. Rev. \textbf{92} (1953) 1145] and {\color{black} are consistent with results for} an undeformed solid subject to a residual stress [Man and Lu,  J. Elasticity \textbf{17} (1987) 159].  These formulas provide a basis for acoustic evaluation of the second- and  third-order elasticity constants and of the residual stresses.  The results are further illustrated in respect of a prototype model of nonlinear elasticity with initial stress, allowing for both finite deformation and nonlinear dependence on the initial stress.

\end{abstract}


\newpage

%++++++++++++++++++++++++++++++++++++++

%%%%%%%%%%

\section{Introduction}

%%%%%%%%%%%

The presence of initial stresses in solid materials can have a substantial effect on their subsequent response to applied loads that is very different from the corresponding response in the absence of initial stresses.  In geophysics, for example, the high stress developed below the Earth's surface due to gravity has a strong influence on the propagation speed of elastic waves, while in soft biological tissues initial (or residual) stresses in artery walls ensure that the circumferential stress distribution through the thickness of the artery wall is close to uniform at typical physiological blood pressures.  Initial stresses may arise, for example, from applied loads, as in the case of gravity, processes of growth and development in living tissue or, in the case of engineering components, from the manufacturing process, either by design to improve the performance of the component or unintentionally (which can lead to defective behaviour).  Here we use the term \emph{initial stress} in its broadest sense, irrespective of how the stress develops.  This includes situations where the stress is due to an applied load leading to an accompanying finite deformation from an
unstressed configuration, in which case the term \emph{prestress} is commonly used, and situations in which the initial stress arises from some other process, such as manufacturing or growth, and is present in the absence of applied loads.  If an initial stress is present in the absence of applied loads (body forces and surface tractions) it is referred to as \emph{residual stress}, as in the definition adopted by \citet{Hoge85}.

In the context of finite deformation elasticity theory much motivation for the study of initial stresses, more especially residual stresses, comes from soft tissue biomechanics, and for a recent discussion of residual stresses in artery walls we refer to \citet{Holz10} and references therein.
The development of constitutive laws for residually stressed materials has been a focus of much of the work of Hoger, as exemplified by \citet{Hoge85,Hoge86,Hoge93} and \citet{John93}; see also the recent paper by \citet{Sara08}.  In the context of acoustoelasticity the work of \cite{Man87} was based on the developments of Hoger, and it also relates to the much earlier work of Biot.
Biot's work was developed within the geophysical context and he was particularly concerned with the effect of initial stress on the propagation of small amplitude elastic waves.
He developed a static theory of small deformations influenced by initial stress \citep{Biot39}, followed by a corresponding theory for wave propagation \citep{Biot40}, all of this work being conveniently collected in his monograph \citep{Biot65}.

Whatever the source of the initial stress and whether or not it is accompanied by a finite deformation, of particular interest in many applications is the effect that the initial stress has on small deformations (static or time-dependent), specifically deformations \emph{linearized} relative to the initially-stressed state and often referred to as \emph{incremental} deformations or motions.
If there is an accompanying finite deformation then the theory is often referred to as the theory of small deformations superimposed on large deformations.
{\color{black}This theory requires knowledge of the associated elasticity tensor, which in general depends on any finite deformation present and on the initial stress.}
Thus, part of the purpose of the present work is to obtain the general form of the elasticity tensor for an initially stressed material that is subject to a finite deformation where the initial stress is not itself associated with an initial finite deformation. We shall be concerned primarily with the case in which the material possesses no intrinsic anisotropy so that any anisotropy arises solely from the presence of initial stress, and we show how the components of the elasticity tensor relate to those of the elasticity coefficients in Biot's isotropic theory.
In the present paper we use the elasticity tensor to examine the effect of initial stress on the propagation of small amplitude elastic waves for both finitely deformed and unstrained materials, with reference to the associated strong ellipticity condition.

In Section \ref{sec2},  the basic equations for a (hyper)elastic material subject to initial stress are summarized together with the equations governing equilibrium of the initially stressed and finitely deformed initially stressed configurations.  The equations of incremental motions superimposed on an initially stressed and finitely deformed equilibrium configuration are then given in Section \ref{sec3}, wherein the elasticity tensor is derived and its general symmetry properties recorded.  In Section \ref{sec4} we then specialize the general theory and focus on the development of the constitutive law for an initially stressed material that has no intrinsic anisotropy such as might be associated with preferred directions, for example, and any anisotropy in its response relative to the initially stressed configuration is entirely due to the initial stress.
The constitutive law of the material is based on a strain-energy function (defined per unit reference volume) that depends on the combined invariants of the right Cauchy--Green deformation tensor and  the initial stress tensor.  For a compressible material there are 10 such independent invariants in the general three-dimensional case, a number which reduces to 9 for an incompressible material.  Expressions for the Cauchy stress and nominal stress tensors and the elasticity tensor are given in general forms for both compressible and incompressible materials and then specialized for specific applications by reducing the number of invariants involved while ensuring that the effects of initial stress are adequately accounted for.  Some details of the calculations are relegated to Appendix A for convenience of reference.

In Section \ref{sec5}, the equations of motion are specialized in order to study the effect of initial stress on the speed of infinitesimal homogeneous plane waves.
First, we give results for the situation in which there is initial stress but no finite deformation.  {\color{black}It is noted, in particular, that the wave speed depends in a nonlinear fashion on the initial stress.}  Results are compared with those arising from Biot's isotropic theory based on connections between the components of the elasticity tensor used here and those used in Biot's theory.  The general connection between these components, which was derived in a recent paper \citep{Ogde11}, is noted here for reference.  {\color{black}Also, within the present framework, by considering a linear elastic material, we confirm a formula given by \cite{Man87} in the general linear theory, which they attribute to Biot, concerning the effect of initial stress on the speed of homogeneous plane waves and the acoustoelastic effect.}  Next, for the Murnaghan form of strain-energy function (\citealp{Murn}), which is appropriate for second-order elastic deformations, and specializing to the case of a material without initial stress, we recover the results of \cite{HuKe53} concerned with the second-order correction for the speeds of longitudinal and transverse waves in an isotropic elastic material.  The final illustration introduces a prototype nonlinear model of an elastic material with initial stress that allows for both finite deformation and nonlinear initial stress.

%%%%%%%%%%%

\section{Basic equations for an elastic solid with initial stress\label{sec2}}

%%%%%%%%%%%

Consider an elastic body in some \emph{reference configuration}, which we denote by  $\mathcal{B}_r$.  Let $\mathbf{X}$ be the position vector of a material point in  $\mathcal{B}_r$.  Any subsequent deformation of the body is measured from  $\mathcal{B}_r$ and we assume that there is an initial (Cauchy) stress $\bm\tau$ in this configuration.  This initial stress is symmetric (rotational balance in the absence of intrinsic couple stresses) and satisfies the equilibrium equation
\begin{equation}
\Div\bm\tau=\mathbf{0},\label{equil-tau}
\end{equation}
in the absence of body forces, where $\Div$ denotes the divergence operator with respect to $\mathcal{B}_r$. If the traction on the boundary $\partial\mathcal{B}_r$ of $\mathcal{B}_r$ vanishes pointwise then $\bm\tau$ is referred to as a \emph{residual stress}, and it is necessarily non-uniform \citep{Hoge85,Ogde03}.  If the traction is not zero then the initial stress may or may not be accompanied by some prior deformation required to reach the configuration $\mathcal{B}_r$ from a completely unstressed configuration.  Here we shall not be concerned with how the initial stress is produced.

Suppose now that the body is deformed quasi-statically into a new configuration $\mathcal{B}$ with boundary $\partial\mathcal{B}$ so that the material point $\mathbf{X}$ takes up the new position $\mathbf{x}$ given by $\mathbf{x}=\bm\chi(\mathbf{X})$, where the vector function $\bm\chi$ defines the deformation for $\mathbf{X}\in\mathcal{B}_r$.  The so-called \emph{deformation} $\bm\chi$ is required to be a bijection and to possess appropriate regularity properties, which need not be made explicit here.  The deformation gradient tensor, denoted $\mathbf{F}$, is defined by $\mathbf{F}=\Grad\bm\chi$, where $\Grad$ is the gradient operator with respect to $\mathcal{B}_r$, and the left and right Cauchy--Green deformation tensors are defined by
\begin{equation}
\mathbf{B}=\mathbf{FF}^\mathrm{T},\quad \mathbf{C}=\mathbf{F}^\mathrm{T}\mathbf{F},
\end{equation}
respectively.

Let $\bm\sigma$ and $\mathbf{S}$ denote the Cauchy stress tensor and the nominal stress tensor, respectively, in the configuration $\mathcal{B}$.  For equilibrium in the absence of body forces $\bm\sigma$ and $\mathbf{S}$ satisfy the equations
\begin{equation}
\div\bm\sigma=\mathbf{0},\quad \Div\mathbf{S}=\mathbf{0},
\end{equation}
and we note the standard connection $\bm\sigma=(\det\mathbf{F})^{-1}\mathbf{FS}$.  For a material without couple stresses, $\bm\sigma$ is symmetric and hence we have
\begin{equation}
\mathbf{FS}=\mathbf{S}^\mathrm{T}\mathbf{F}^\mathrm{T}.\label{symmetry}
\end{equation}

For an elastic material we consider a strain-energy function $W$ defined per unit volume in $\mathcal{B}_r$.
This function depends on the deformation gradient $\mathbf{F}$ and the initial stress $\bm\tau$.  If $\bm\tau$ depends on $\mathbf{X}$, as would be the case for a residually-stressed material, then the material is necessarily inhomogeneous, but if $\bm\tau$ is independent of $\mathbf{X}$ the material is homogeneous unless its properties depend separately on $\mathbf{X}$.  In either case we make the dependence on $\bm\tau$ explicit and write
\begin{equation}
W=W(\mathbf{F},\bm\tau).
\end{equation}
Of course, by objectivity, $W$ depends on $\mathbf{F}$ only through $\mathbf{C}=\mathbf{F}^\mathrm{T}\mathbf{F}$, but otherwise this form of $W$ is completely general and no material symmetry is invoked.
Note, however, that in general, i.e. if $\bm\tau$ is not a hydrostatic compression or tension, the presence of $\bm\tau$ induces some anisotropy in the material, even if the material has no intrinsic anisotropy.
Thus, $\bm\tau$ has an effect on the constitutive law analogous to that of a structure tensor in anisotropic elasticity.

For a material not subject to any internal constraints, the nominal and Cauchy stresses are given by
\begin{equation}
\mathbf{S}=\frac{\partial W}{\partial\mathbf{F}}(\mathbf{F},\bm\tau), \quad \bm\sigma=J^{-1}\mathbf{FS}=J^{-1}\mathbf{F}\frac{\partial W}{\partial\mathbf{F}}(\mathbf{F},\bm\tau),\label{stress-comp}
\end{equation}
respectively, where $J=\det\mathbf{F}>0$.  When evaluated in $\mathcal{B}_r$, these give the connection
\begin{equation}
\bm\tau=\frac{\partial W}{\partial\mathbf{F}}(\mathbf{I},\bm\tau),\label{stress-comp-ref}
\end{equation}
where $\mathbf{I}$ is the identity tensor.

For an \emph{incompressible material}, the internal constraint
\begin{equation}
J\equiv\det\mathbf{F}=1
\end{equation}
must hold for all deformations and the counterpart of \eqref{stress-comp} in this case is
\begin{equation}
\mathbf{S}=\frac{\partial W}{\partial\mathbf{F}}(\mathbf{F},\bm\tau)-p\mathbf{F}^{-1}, \quad \bm\sigma=\mathbf{FS}=\mathbf{F}\frac{\partial W}{\partial\mathbf{F}}(\mathbf{F},\bm\tau)-p\mathbf{I},\label{stress-incomp}
\end{equation}
where $p$ is a Lagrange multiplier associated with the constraint. When evaluated in the residually stressed configuration, these both reduce to
\begin{equation}
\bm\tau=\frac{\partial W}{\partial\mathbf{F}}(\mathbf{I},\bm\tau)-p^{(r)}\mathbf{I},\label{stress-incomp-ref}
\end{equation}
where $p^{(r)}$ is the value of $p$ in $\mathcal{B}_r$.

%%%%%%%%%%%%%%%%%%%%%

\section{Incremental motion superimposed on an initially stressed configuration subject to finite deformation
\label{sec3}}

%%%%%%%%%%%%%%%%%%%%%

Superimposed on the equilibrium configuration $\mathcal{B}$ defined by $\mathbf{x}=\bm\chi(\mathbf{X})$, we now consider an incremental motion $\mathbf{\dot{x}}(\mathbf{X},t)$, where $t$ is time.
Here and in the following a superposed dot indicates an incremental quantity, increments are consider `small', and the resulting incremental equations are linearized in the increments.
Thus, $\mathbf{\dot{x}}$ represents the displacement from $\mathbf{x}$, and we shall also express it in Eulerian form by writing the displacement vector as a function of $\mathbf{x}$ and $t$, namely $\mathbf{u}=\mathbf{u}(\mathbf{x},t)$.  The corresponding increment in the deformation gradient, $\mathbf{\dot{F}}$, is expressible as
\begin{equation}
\mathbf{\dot{F}}=\mathbf{LF},\label{Fdot}
\end{equation}
where $\mathbf{L}=\grad\mathbf{u}$ is the displacement gradient.

The (linearized) incremental nominal stress takes the forms
\begin{equation}
\mathbf{\dot{S}}=\left\{
\begin{array}{l}
\bm{\mathcal{A}}\mathbf{\dot{F}},\\[0.1cm]
\bm{\mathcal{A}}\mathbf{\dot{F}}+p\mathbf{F}^{-1}\mathbf{\dot{F}}\mathbf{F}^{-1}-\dot{p}\mathbf{F}^{-1},\label{incrementalS}
\end{array}
\right.
\end{equation}
for unconstrained and incompressible materials, respectively, where
\begin{equation}
\bm{\mathcal{A}}
= \frac{\partial^2W}{\partial\mathbf{F}\partial\mathbf{F}},
\qquad
\mathcal{A}_{\alpha i\beta j}=\frac{\partial^2W}{\partial F_{i\alpha}\partial F_{j\beta}},\label{mathcalA-def}
\end{equation}
is the \emph{elasticity tensor} and, in component form, $\bm{\mathcal{A}}\mathbf{\dot{F}}\equiv \mathcal{A}_{\alpha i\beta j}\dot{F}_{j\beta}$ defines the product used in the above.  Here and henceforth the standard summation convention for repeated indices applies. We note in passing that \eqref{incrementalS} is applicable even if a strain-energy function is not assumed, in which case the major symmetry $i\alpha \leftrightarrow j\beta$ from \eqref{mathcalA-def} is lost; moreover, the theory could be considered more general if the explicit dependence on structure tensors such as $\bm\tau$ is omitted.  The generality considered here, however, is sufficient for our purpose.

For an incompressible material the incremental incompressibility can be written in either of the forms
\begin{equation}
\tr(\mathbf{\dot{F}}\mathbf{F}^{-1})=\tr\mathbf{L}=0, \qquad \div\mathbf{u}=0.
\end{equation}

In the absence of body forces the incremental motion is governed by the equation
\begin{eqnarray}
\Div\mathbf{\dot{S}}=\rho_r\mathbf{x}_{,tt},\label{motion1}
\end{eqnarray}
where $\rho_r$ is the mass density in $\mathcal{B}_r$ and a subscript $t$ following a comma signifies the material time derivative, i.e. the time derivative at fixed $\mathbf{X}$, so that $\mathbf{x}_{,t}=\mathbf{u}_{,t}$ is the particle velocity and $\mathbf{x}_{,tt}=\mathbf{u}_{,tt}$ the acceleration.  The incremental counterpart of the rotational balance equation \eqref{symmetry} is
\begin{equation}
\mathbf{F}\mathbf{\dot{S}}+\mathbf{\dot{F}}\mathbf{S}=\mathbf{\dot{S}}^\mathrm{T}\mathbf{F}^\mathrm{T}+\mathbf{S}^\mathrm{T}\mathbf{\dot{F}}^\mathrm{T},
\end{equation}
or, on use of \eqref{stress-comp}$_2$ and \eqref{Fdot},
\begin{equation}
\mathbf{F}\mathbf{\dot{S}}+J\mathbf{L}\bm\sigma=\mathbf{\dot{S}}^\mathrm{T}\mathbf{F}^\mathrm{T}+J\bm\sigma\mathbf{L}^\mathrm{T}.\label{symmetry2}
\end{equation}
On introducing the `push forward' $\mathbf{\dot{S}}_0$ of $\mathbf{\dot{S}}$, defined by $\mathbf{\dot{S}}_0=J^{-1}\mathbf{F}\mathbf{\dot{S}}$, and the corresponding push forward $\bm{\mathcal{A}}_0$ of the elasticity tensor  such that $\mathbf{\dot{S}}_0=\bm{\mathcal{A}}_0\mathbf{L}$ ($\mbox{or\ } \bm{\mathcal{A}}_0\mathbf{L} +\,p\mathbf{L}-\dot{p}\mathbf{I}$ in the case of an incompressible material) we may express
\eqref{symmetry2} in the form
\begin{equation}
\bm{\mathcal{A}}_0\mathbf{L}+\mathbf{L}\bm\sigma=(\bm{\mathcal{A}}_0\mathbf{L})^\mathrm{T}+\bm\sigma\mathbf{L}^\mathrm{T}\label{symmetry3}
\end{equation}
for an unconstrained material, and as
\begin{equation}
\bm{\mathcal{A}}_0\mathbf{L}+\mathbf{L}(\bm\sigma+p\mathbf{I})=(\bm{\mathcal{A}}_0\mathbf{L})^\mathrm{T}+(\bm\sigma+p\mathbf{I})\mathbf{L}^\mathrm{T},\label{symmetry4}
\end{equation}
for an incompressible material.  In component form $\bm{\mathcal{A}}_0$ is related to $\bm{\mathcal{A}}$ via
\begin{equation}
J\mathcal{A}_{0piqj}=F_{p\alpha}F_{q\beta}\mathcal{A}_{\alpha i\beta j},\label{AA0connect}
\end{equation}
with $J=1$ in the incompressible case.

For details of the background on the theory of incremental deformations superimposed on a finite deformation we refer to \citet{Ogde84,Ogde07}, for example.

From \eqref{symmetry3} we deduce that $\tr[(\bm{\mathcal{A}}_0\mathbf{L}+\mathbf{L}\bm\sigma)\mathbf{W}]=0$ for any skew symmetric second-order tensor $\mathbf{W}$, and since this and the corresponding result from \eqref{symmetry4} holds for arbitrary $\mathbf{L}$ we deduce further that
\begin{equation}
\bm{\mathcal{A}}_0\mathbf{W}=-\bm\sigma\mathbf{W},\quad \bm{\mathcal{A}}_0\mathbf{W}=-(\bm\sigma+p\mathbf{I})\mathbf{W},\label{HogerW-result}
\end{equation}
for an arbitrary skew symmetric $\mathbf{W}$ for unconstrained and incompressible materials, respectively. A formula equivalent to \eqref{HogerW-result}$_1$ was derived by \citet{Hoge86} by a different method and stated explicitly as the formula (2.2.2) in the latter paper for a residually-stressed but undeformed configuration.  Note that the derivation herein does not require the existence of a strain-energy function. Another result in \citet{Hoge86} is recovered by choosing $\mathbf{L}$ to be symmetric, which leads to
\begin{equation}
\frac{1}{2}[\bm{\mathcal{A}}_0\mathbf{E}-(\bm{\mathcal{A}}_0\mathbf{E})^\mathrm{T}]=\frac{1}{2}(\bm\sigma\mathbf{E}-\mathbf{E}\bm\sigma),
\end{equation}
for any symmetric $\mathbf{E}$.
This relation applies for both unconstrained and incompressible materials and corresponds to the formula (2.2.3) in \citet{Hoge86}, the right-hand side of which contains a sign error.

Note that as well as possessing the major symmetry $\mathcal{A}_{0piqj}=\mathcal{A}_{0qjpi}$, which follows from  \eqref{mathcalA-def} and \eqref{AA0connect}, $\bm{\mathcal{A}}_0$ has the property
\begin{eqnarray}
\mathcal{A}_{0piqj}+\delta_{jp}\sigma_{iq}=\mathcal{A}_{0ipqj}+\delta_{ij}\sigma_{pq},
\end{eqnarray}
for a compressible material, and
\begin{eqnarray}
\mathcal{A}_{0piqj}+\delta_{jp}(\sigma_{iq}+p\delta_{iq})=\mathcal{A}_{0ipqj}
+\delta_{ij}(\sigma_{pq}+p\delta_{pq}),
\end{eqnarray}
for an incompressible material.  These formulas follow from \eqref{symmetry3} and \eqref{symmetry4}, respectively.  Corresponding formulas for a pre-stressed material in the absence of residual stress were given by \citet{Chad71} and \citet{Chad97}.

In Eulerian form, i.e. in terms of $\mathbf{\dot{S}}_0$ and $\mathbf{u}$, the incremental equation of motion \eqref{motion1} becomes
\begin{equation}
\div\mathbf{\dot{S}}_0=\rho\mathbf{u}_{,tt},
\end{equation}
where $\rho=\rho_rJ^{-1}$ is the mass density in $\mathcal{B}$.  The (Cartesian) component form of this equation may be written, for an unconstrained material, as
\begin{equation}
(\mathcal{A}_{0piqj}u_{j,q})_{,p}=\rho u_{i,tt},\label{motioncomp}
\end{equation}
and for an incompressible material as
\begin{equation}
(\mathcal{A}_{0piqj}u_{j,q})_{,p}-\dot{p}_{,i}+p_{,j}u_{j,i}=\rho_r u_{i,tt},\quad \mbox{with}\quad u_{i,i}=0.\label{motionincomp}
\end{equation}

Suppose now that the material properties are homogeneous, implying, in particular, that the initial stress $\bm\tau$ is uniform, and that the deformation in $\mathcal{B}$ is homogeneous.  Then, $\bm{\mathcal{A}}$, $\bm{\mathcal{A}}_0$ and $p$ are constant, and the equations of motion \eqref{motioncomp} and \eqref{motionincomp} reduce to
\begin{eqnarray}
\mathcal{A}_{0piqj}u_{j,pq}=\rho u_{i,tt},
\end{eqnarray}
and
\begin{eqnarray}
\mathcal{A}_{0piqj}u_{j,pq}-\dot{p}_{,i}=\rho u_{i,tt},\quad u_{i,i}=0,\label{motioninc}
\end{eqnarray}
respectively, and we note that $\rho=\rho_r$ in the latter case.

As already indicated, we are considering $\bm{\mathcal{A}}$ to depend on the deformation through $\mathbf{F}$, on the initial stress $\bm\tau$, and on any material symmetry present in the configuration $\mathcal{B}_r$.  In what follows we consider the intrinsic properties of the material to be isotropic relative to $\mathcal{B}_r$ so that the only source of anisotropy is the initial stress.  In order to account for the initial stress in the expressions for the stress and elasticity tensors we base the development in the following sections on scalar invariants involving the right Cauchy--Green deformation tensor $\mathbf{C}$ and the initial stress tensor $\bm\tau$.

%%%%%%%%%%%%%%%%%%%%%

\section{Invariant-based formulation \label{sec4}}

%%%%%%%%%%%%%%%%%%%%%

%%%%%%%%%%%%%%%%%%%%%%%%%%%%

\subsection{Invariants of $\mathbf{C}$ and $\bm\tau$}

%%%%%%%%%%%%%%%%%%%%%%%%%%%%

The strain-energy function $W$ of an initially stressed material depends on $\bm\tau$ as well as the deformation gradient $\mathbf{F}$, and, by objectivity, it depends on $\mathbf{F}$ only through $\mathbf{C}=\mathbf{F}^\mathrm{T}\mathbf{F}$.  For a material that is isotropic relative to $\mathcal{B}_r$ in the absence of initial stress, $W$ can be treated as a function of three independent invariants of $\mathbf{C}$, which are commonly taken to be the principal invariants defined by
\begin{equation}
I_1=\tr\mathbf{C},\quad I_2=\tfrac{1}{2}[(\tr\mathbf{C})^2-\tr(\mathbf{C}^2)],\quad I_3=\det\mathbf{C}.\label{invs1}
\end{equation}
In the configuration $\mathcal{B}_r$ these reduce to $I_1=I_2=3,\, I_3=1$.  If there is an initial stress $\bm\tau$ in $\mathcal{B}_r$ then, in general, the material response relative to $\mathcal{B}_r$ depends also on the invariants of  $\bm\tau$ and the combined invariants of  $\bm\tau$ and $\mathbf{C}$.
A possible set of independent invariants of $\bm\tau$, including those that depend on $\mathbf{C}$, is
\begin{eqnarray}
\tr\bm\tau,\quad \tr(\bm\tau^2),\quad \tr(\bm\tau^3),\quad \tr(\bm\tau\mathbf{C}),\quad \tr(\bm\tau\mathbf{C}^2),\quad \tr(\bm\tau^2\mathbf{C}),\quad \tr(\bm\tau^2\mathbf{C}^2).\label{invs3}
\end{eqnarray}
Together, there are thus at most 10 independent invariants of $\mathbf{C}$ and $\bm\tau$ in general, and for an incompressible material we have $I_3=1$ and hence there are at most 9 independent invariants in this case.
The number of independent invariants is reduced when the dimension of the considered problem is reduced from three to two, as for a plane strain setting, for example.  In the reference configuration $\mathcal{B}_r$ the fourth and fifth invariants listed in \eqref{invs3} reduce to the first, and the sixth and seventh to the second.  For relevant background on invariants of tensors we refer to \citet{Spen71} and \citet{Zhen94}.  For subsequent reference we introduce the notation
\begin{equation}
I_6=\tr(\bm\tau\mathbf{C}),\quad I_7= \tr(\bm\tau\mathbf{C}^2),\quad I_8=\tr(\bm\tau^2\mathbf{C}),\quad I_9=\tr(\bm\tau^2\mathbf{C}^2),
\end{equation}
for the invariants that depend on both $\mathbf{C}$ and $\bm\tau$, noting that the notations $I_4$ and $I_5$ are not used since these tend to be associated with a preferred direction in a transversely isotropic material.

Although not immediately obvious, it is worth noting that invariants such as $\tr(\mathbf{C}\bm\tau\mathbf{C}\bm\tau)$ and $\tr(\mathbf{C}\bm\tau\mathbf{C}^2\bm\tau)$ may be expressed in terms of the invariants listed above.  This may be shown by first applying the Cayley--Hamilton theorem to $\mathbf{C}+\lambda\bm\tau$ for an arbitrary scalar $\lambda$.  The coefficient of $\lambda$ yields the identity
\begin{eqnarray}
\mathbf{C}^2\bm\tau+\mathbf{C}\bm\tau\mathbf{C}+\bm\tau\mathbf{C}^2&-&I_1(\mathbf{C}\bm\tau+\bm\tau\mathbf{C})-(\tr\bm\tau)\mathbf{C}^2
\notag\\[1ex]
 &-&[I_1(\tr\bm\tau)-I_6]\mathbf{C}+I_2\bm\tau-[I_2(\tr\bm\tau)-I_1I_6+I_7]\mathbf{I}=\mathbf{0}.
\end{eqnarray}
Then, by multiplying by $\bm\tau$, for example, and then taking the trace of the result leads to
\begin{equation}
\tr(\mathbf{C}\bm\tau\mathbf{C}\bm\tau)=2I_1I_8-2I_9-2I_1I_6(\tr\bm\tau)+2I_7(\tr\bm\tau)+I_6^2+I_2[(\tr\bm\tau)^2-\tr(\bm\tau^2)],
\end{equation}
which shows how $\tr(\mathbf{C}\bm\tau\mathbf{C}\bm\tau)$ depends on the other invariants.

The expressions for the stress and elasticity tensors given in \eqref{stress-comp}, \eqref{stress-incomp} and \eqref{mathcalA-def}  require the calculation of
\begin{eqnarray}
\frac{\partial W}{\partial\mathbf{F}}=\sum_{i\,\in\,\mathcal{I}} W_i\frac{\partial I_i}{\partial \mathbf{F}},\label{partialW}
\end{eqnarray}
and
\begin{eqnarray}
\frac{\partial^2 W}{\partial\mathbf{F}\partial\mathbf{F}}=\sum_{i\,\in\,\mathcal{I}} W_i\frac{\partial^2 I_i}{\partial \mathbf{F}\partial \mathbf{F}}
+\sum_{i,j\,\in\,\mathcal{I} }W_{ij}\frac{\partial I_i}{\partial\mathbf{F}}\otimes\frac{\partial I_j}{\partial\mathbf{F}},\label{partial2W}
\end{eqnarray}
where we have used the shorthand notations $W_i=\partial W/\partial I_i,\, W_{ij}=\partial^2 W/\partial I_i\partial I_j,\, i,j\in\mathcal{I}$, and $\mathcal{I}$ is the index set $\{1,2,3,6,7,8,9\}$ (or $\{1,2,6,7,8,9\}$ in the case of an incompressible material).  Although their derivatives with respect to $\mathbf{F}$ vanish and do not appear in the above expressions, the invariants $\tr\bm\tau$, $\tr(\bm\tau^2)$ and $\tr(\bm\tau^3)$ may nevertheless be included in the functional dependence of $W$.  In the following we give explicit expressions for the stress tensors and the elasticity tensor based on these summations.

%%%%%%%%%%%%%%%

\subsection{Stress tensors}

%%%%%%%%%%%%%%%

For an unconstrained material the strain-energy function $W$ depends on the seven deform\-ation dependent invariants $I_1,I_2,I_3,I_6,\dots,I_9$ together, in general, with $\tr\bm\tau$, $\tr(\bm\tau^2)$ and $\tr(\bm\tau^3)$.  From \eqref{stress-comp} and \eqref{partialW} the nominal and Cauchy stresses are given by
\begin{equation}
\mathbf{S}=\frac{\partial W}{\partial \mathbf{F}}=\sum_{i\,\in\,\mathcal{I}}W_i\frac{\partial I_i}{\partial \mathbf{F}},\qquad \bm\sigma=J^{-1}\mathbf{FS}.
\end{equation}
The corresponding expressions for an incompressible material are
\begin{equation}
\mathbf{S}=\frac{\partial W}{\partial \mathbf{F}}-p\mathbf{F}^{-1}=\sum_{i\,\in\,\mathcal{I}}W_i\frac{\partial I_i}{\partial \mathbf{F}}-p\mathbf{F}^{-1},\qquad
\bm\sigma=\mathbf{FS}.
\end{equation}
The required expressions for $\partial I_i/\partial \mathbf{F}$ are listed for convenience in Appendix A.  In particular,  these enable the Cauchy stress to be expanded, so that
\begin{eqnarray}
J\bm\sigma &=& 2 W_1 \mathbf{B} + 2 W_2 (I_1 \mathbf{B} - \mathbf{B}^2) + 2I_3W_3 \mathbf{I} + 2 W_6 \bm\Sigma\notag
\\[1ex]
 &+& 2 W_7 (\bm\Sigma\mathbf{B} + \mathbf{B}\bm\Sigma)
 + 2W_8 \bm\Sigma \mathbf{B}^{-1} \bm\Sigma
 + 2 W_9(\bm\Sigma\mathbf{B}^{-1} \bm\Sigma\mathbf{B}
   + \mathbf{B} \bm\Sigma \mathbf{B}^{-1} \bm\Sigma),\label{comp-sigma}
\end{eqnarray}
where the notation $\bm\Sigma=\mathbf{F}\bm\tau\mathbf{F}^\mathrm{T}$ has been introduced and we recall that $\mathbf{B}=\mathbf{FF}^\mathrm{T}$ is the left Cauchy--Green tensor.  For an incompressible material \eqref{comp-sigma} is replaced by
\begin{eqnarray}
\bm\sigma &=&
2 W_1 \mathbf{B} + 2 W_2 (I_1 \mathbf{B} - \mathbf{B}^2) -p \mathbf{I} + 2 W_6 \bm\Sigma\notag
\\[1ex]
 &+& 2 W_7 (\bm\Sigma\mathbf{B} + \mathbf{B}\bm\Sigma)
 + 2W_8 \bm\Sigma \mathbf{B}^{-1} \bm\Sigma
 + 2 W_9(\bm\Sigma\mathbf{B}^{-1} \bm\Sigma\mathbf{B}
   + \mathbf{B} \bm\Sigma \mathbf{B}^{-1} \bm\Sigma),\label{incomp-sigma}
\end{eqnarray}
with $I_3\equiv 1$.

If we evaluate \eqref{comp-sigma} in the reference configuration, it reduces to the appropriate specialization of \eqref{stress-comp-ref}, namely
\begin{equation}
\bm\tau = 2(W_1 + 2W_2 +W_3)\mathbf{I} + 2(W_6 + 2 W_7)\bm\tau + 2(W_8 + 2 W_9)\bm\tau^{2},\label{sigma-in-ref-comp}
\end{equation}
where all $W_i,\,i\in\mathcal{I}$, are evaluated for $I_1=I_2=3,\,I_3=1$, $I_6=I_7=\tr\bm\tau$, $I_8=I_9=\tr(\bm\tau^2)$.
This indicates that we should set
\begin{equation}
W_1 + 2 W_2 +W_3 = 0, \quad 2(W_6 + 2 W_7) = 1, \quad W_8 + 2 W_9 = 0\label{ref-conds-comp}
\end{equation}
there.
In fact, if we require \eqref{sigma-in-ref-comp} to hold \emph{for all} $\bm\tau$ then the conditions in \eqref{ref-conds-comp} necessarily follow.
The corresponding reduction for an incompressible material yields
\begin{equation}
\bm\tau = (2W_1 + 4W_2 -p^{(r)})\mathbf{I} + 2(W_6 + 2 W_7)\bm\tau + 2(W_8 + 2 W_9)\bm\tau^{2},\label{sigma-in-ref-incomp}
\end{equation}
which specializes \eqref{stress-incomp-ref},
and the counterpart of \eqref{ref-conds-comp} is
\begin{equation}
2 W_1 + 4 W_2 - p^{(r)} = 0, \quad 2(W_6 + 2 W_7) = 1, \quad W_8 + 2 W_9 = 0,\label{ref-conds-incomp}
\end{equation}
evaluated in the reference configuration.

%%%%%%%%%%%%%%%%%%%%%%%%

\subsection{The elasticity tensor\label{sec4-3}}

%%%%%%%%%%%%%%%%%%%%%%%%

Next, we note that the elasticity tensor $\bm{\mathcal{A}}$ is given by
\begin{eqnarray}
\bm{\mathcal{A}}=\frac{\partial^2W}{\partial\mathbf{F}\partial\mathbf{F}}=\sum_{i\,\in\,\mathcal{I}}W_i\frac{\partial^2I_i}{\partial\mathbf{F}\partial\mathbf{F}}+\sum_{i,j\,\in\,\mathcal{I}}W_{ij}\frac{\partial I_i}{\partial\mathbf{F}}\otimes\frac{\partial I_j}{\partial\mathbf{F}}.
\end{eqnarray}
This requires expressions for the second derivatives of the invariants, which are collected together for reference in Appendix A in component form.  Here we give the component form of $\bm{\mathcal{A}}_0$:
\begin{eqnarray}
J\mathcal{A}_{0piqj}&=&2(W_1+I_1W_2)B_{pq}\delta_{ij}+2W_2[2B_{pi}B_{qj}-B_{iq}B_{jp}-B_{pr}B_{rq}\delta_{ij}-B_{pq}B_{ij}]\notag\\[0.1cm]
&+&2I_3W_3(2\delta_{ip}\delta_{jq}-\delta_{iq}\delta_{jp})+2W_6\Sigma_{pq}\delta_{ij}+2W_7[\Sigma_{pq}B_{ij}+\Sigma_{pr} B_{rq}\delta_{ij}+B_{pr}\Sigma_{rq}\delta_{ij}\notag\\[0.1cm]
&+&\Sigma_{ij}B_{pq}+\Sigma_{pj}B_{iq}+\Sigma_{iq}B_{pj}]+4W_{11}B_{ip}B_{jq}\notag\\[0.1cm]
&+&4W_{22}(I_1 B_{ip}-B_{ir}B_{rp})(I_1 B_{jq}-B_{js}B_{sq})\notag\\[0.1cm]
&+&4I_3^2W_{33}\delta_{ip}\delta_{jq}+4W_{12}(2I_1B_{ip}B_{jq}-B_{ip}B_{jr}B_{rq}-B_{jq}B_{ir}B_{rp})\notag\\[0.1cm]
&+&4I_3W_{13}(B_{ip}\delta_{jq}+B_{jq}\delta_{ip})+4I_3W_{23}[I_1(B_{ip}\delta_{jq}+B_{jq}\delta_{ip})-\delta_{ip}B_{jr}B_{rq}-\delta_{jq}B_{ir}B_{rp}]\notag\\[0.1cm]
&+&4W_{16}(B_{ip}\Sigma_{jq}+B_{jq}\Sigma_{ip})+4W_{17}[B_{ip}(\Sigma_{jr} B_{rq}+B_{jr}\Sigma_{rq})+B_{jq}(\Sigma_{ir} B_{rp}+B_{ir}\Sigma_{rp})]\notag\\[0.1cm]
&+&4W_{26}[(I_1B_{ip}-B_{ir}B_{rp})\Sigma_{jq}+(I_1B_{jq}-B_{jr}B_{rq})\Sigma_{ip}]\notag\\[0.1cm]
&+&4W_{27}[(I_1B_{ip}-B_{ir}B_{rp})(\Sigma_{js} B_{sq}+B_{js}\Sigma_{sq})+(I_1B_{jq}-B_{jr}B_{rq})(\Sigma_{is} B_{sp}+B_{is}\Sigma_{sp})]\notag\\[0.1cm]
&+&4I_3W_{36}(\delta_{ip}\Sigma_{jq}+\delta_{jq}\Sigma_{ip})+4I_3W_{37}[\delta_{ip}(\Sigma_{jr} B_{rq}+B_{jr}\Sigma_{rq})+\delta_{jq}(\Sigma_{ir} B_{rp}+B_{ir}\Sigma_{rp})]\notag\\[0.1cm]
&+&4W_{66}\Sigma_{ip}\Sigma_{jq}+4W_{67}[\Sigma_{ip}(\Sigma_{jr} B_{rq}+B_{jr}\Sigma_{rq})+\Sigma_{jq}(\Sigma_{ir} B_{rp}+B_{ir}\Sigma_{rp})]\notag\\[0.1cm]
&+&4W_{77}(\Sigma_{ir} B_{rp}+B_{ir}\Sigma_{rp})(\Sigma_{js} B_{sq}+B_{js}\Sigma_{sq}),\label{mathcalA0-components-comp}
\end{eqnarray}
and, in order to save space, terms involving derivatives with respect to $I_8$ or $I_9$ have not been included.  Note, however, that it is straightforward to include these terms, just by repeating the terms involving derivatives with respect to $I_6$ and $I_7$, replacing the indices 6 and 7 by 8 and 9, respectively, and replacing $\bm\Sigma=\mathbf{F}\bm\tau\mathbf{F}^{\mathrm{T}}$ by $\mathbf{F}\bm\tau^2\mathbf{F}^{\mathrm{T}}$. When there is no initial stress the above formula reduces to the result for a prestressed isotropic elastic solid, as given in, for example, the classical paper by \cite{Haye61}; see also \cite{Toup61} and \cite{True61} for related basic theory.

The corresponding formula for an incompressible
is obtained by setting $J=1$ and $I_3=1$ and omitting the terms involving derivatives of $W$ with respect to $I_3$, and yields
\begin{eqnarray}
\mathcal{A}_{0piqj}&=&2(W_1+I_1W_2)B_{pq}\delta_{ij}+2W_2[2B_{pi}B_{qj}-B_{iq}B_{jp}-B_{pr}B_{rq}\delta_{ij}-B_{pq}B_{ij}]\notag\\[0.1cm]
&+&2W_6\Sigma_{pq}\delta_{ij}+2W_7[\Sigma_{pq}B_{ij}+\Sigma_{pr} B_{rq}\delta_{ij}+B_{pr}\Sigma_{rq}\delta_{ij}+\Sigma_{ij}B_{pq}+\Sigma_{pj}B_{iq}+\Sigma_{iq}B_{pj}]
\notag\\[0.1cm]
&+&4W_{11}B_{ip}B_{jq}+4W_{22}(I_1 B_{ip}-B_{ir}B_{rp})(I_1 B_{jq}-B_{js}B_{sq})\notag\\[0.1cm]
&+&4W_{12}(2I_1B_{ip}B_{jq}-B_{ip}B_{jr}B_{rq}-B_{jq}B_{ir}B_{rp})\notag\\[0.1cm]
&+&4W_{16}(B_{ip}\Sigma_{jq}+B_{jq}\Sigma_{ip})+4W_{17}[B_{ip}(\Sigma_{jr} B_{rq}+B_{jr}\Sigma_{rq})+B_{jq}(\Sigma_{ir} B_{rp}+B_{ir}\Sigma_{rp})]\notag\\[0.1cm]
&+&4W_{26}[(I_1B_{ip}-B_{ir}B_{rp})\Sigma_{jq}+(I_1B_{jq}-B_{jr}B_{rq})\Sigma_{ip}]\notag\\[0.1cm]
&+&4W_{27}[(I_1B_{ip}-B_{ir}B_{rp})(\Sigma_{js} B_{sq}+B_{js}\Sigma_{sq})+(I_1B_{jq}-B_{jr}B_{rq})(\Sigma_{is} B_{sp}+B_{is}\Sigma_{sp})]\notag\\[0.1cm]
&+&4W_{66}\Sigma_{ip}\Sigma_{jq}+4W_{67}[\Sigma_{ip}(\Sigma_{jr} B_{rq}+B_{jr}\Sigma_{rq})+\Sigma_{jq}(\Sigma_{ir} B_{rp}+B_{ir}\Sigma_{rp})]\notag\\[0.1cm]
&+&4W_{77}(\Sigma_{ir} B_{rp}+B_{ir}\Sigma_{rp})(\Sigma_{js} B_{sq}+B_{js}\Sigma_{sq}).\label{mathcalA0-components-incomp}
\end{eqnarray}

When evaluated in the reference configuration, the moduli \eqref{mathcalA0-components-comp}, with the missing terms included, reduce to
\begin{eqnarray}
\mathcal{A}_{0piqj}
&=&
\alpha_1 (\delta_{ij}\delta_{pq}+\delta_{iq}\delta_{jp})
+
\alpha_2 \delta_{ip}\delta_{jq}+\delta_{ij}\tau_{pq}+\beta_1(\delta_{ij}\tau_{pq}+\delta_{pq}\tau_{ij}+\delta_{iq}\tau_{jp}+\delta_{jp}\tau_{iq})\notag
\\[1ex]
&+&
\beta_2(\delta_{ip}\tau_{jq} +\delta_{jq}\tau_{ip})
+
\beta_3\tau_{ip}\tau_{jq}+\gamma_1(\delta_{ij}\tau_{pk}\tau_{kq} + \delta_{pq}\tau_{ik}\tau_{kj} + \delta_{iq}\tau_{jk}\tau_{kp} + \delta_{jp}\tau_{ik}\tau_{kq})\notag
\\[1ex]
&+&
\gamma_2(\delta_{ip}\tau_{jk}\tau_{kq} + \delta_{jq}\tau_{ik}\tau_{kp})+\gamma_3(\tau_{ip}\tau_{jk}\tau_{kq}+\tau_{jq}\tau_{ik}\tau_{kp})
+
\gamma_4\tau_{ik}\tau_{kp}\tau_{jk}\tau_{kq},\label{A0-comp-components-ref}
\end{eqnarray}
where the $\alpha$'s, $\beta$'s, and $\gamma$'s are defined by
\begin{align}
& \alpha_1 = 2(W_1+W_2),
\qquad
\alpha_2 = 4(W_{11}+4W_{12}+4W_{22}+2W_{13}+4W_{23}+W_{33}) - 2\alpha_1,
 \notag \\[0.1cm]
& \beta_1 = 2W_7,
\qquad
\beta_2 = 4(W_{16}+2W_{17}+2W_{26}+4W_{27}+W_{36}+2W_{37}),
 \notag \\[0.1cm]
&\beta_3 = 4(W_{66}+4W_{67}+4W_{77}),
\qquad
 \gamma_1 = 2W_9,
  \notag \\[0.1cm]
&
\gamma_2 = 4(W_{18} + 2W_{19} + 2W_{28}+4W_{29}+W_{38} + 2W_{39}),\notag\\[0.1cm]
&
\gamma_3=4(W_{68}+2W_{69}+2W_{78}+4W_{79}),
\qquad
\gamma_4 = 4(W_{88} + 4W_{89} + 4W_{99}),
\label{alpha--epsilon}
\end{align}
all derivatives being evaluated in the reference configuration and use having been made of the connections \eqref{ref-conds-comp}.
Note that in general the expressions \eqref{alpha--epsilon} may depend on the invariants $\tr\bm\tau$, $\tr(\bm\tau^2)$ and $\tr(\bm\tau^3)$.

It is worth noting here that when referred to axes that coincide with the principal axes of $\bm\tau$, the only non-zero components of \eqref{A0-comp-components-ref} are given by
\begin{align}
&
\mathcal{A}_{0iiii}
= 2\alpha_1+\alpha_2+(1+4\beta_1+2\beta_2)\tau_i+\beta_3\tau_i^2+2(2\gamma_1+\gamma_2)\tau_i^2+2\gamma_3\tau_i^3+\gamma_4\tau_i^4,
\notag \\[0.1cm]
&
\mathcal{A}_{0iijj}
=\alpha_2+\beta_2(\tau_i+\tau_j)+\beta_3\tau_i\tau_j+\gamma_2(\tau_i^2+\tau_j^2)+\gamma_3(\tau_i+\tau_j)\tau_i\tau_j+\gamma_4\tau_i^2\tau_j^2,
\notag \\[0.1cm]
&
\mathcal{A}_{0ijij}
=\alpha_1+\tau_i+\beta_1(\tau_i+\tau_j)+\gamma_1(\tau_i^2+\tau_j^2)=\mathcal{A}_{0ijji}+\tau_i,
\end{align}
where there are no sums on repeated indices, $i\neq j$, and $\tau_i$ ($i=1,2,3$) are the principal values of $\bm\tau$ (in general, 15 non-zero components of $\bm{\mathcal{A}_0}$ in total).  It is interesting, but not surprising, to note that for a prestressed isotropic elastic material the only non-zero components of $\bm{\mathcal{A}}_0$ are also $\mathcal{A}_{0iiii}$,  $\mathcal{A}_{0iijj}$,  $\mathcal{A}_{0ijij}$ and $\mathcal{A}_{0ijji}$,\,$i\neq j$ (see, for example, \citealp{Ogde84}).

For an incompressible material there is a further reduction of \eqref{A0-comp-components-ref}, leading to
\begin{eqnarray}
\mathcal{A}_{0piqj}
&=&
\alpha_1 (\delta_{ij}\delta_{pq}+\delta_{iq}\delta_{jp})+\delta_{ij}\tau_{pq}+
\beta_1(\delta_{ij}\tau_{pq}+\delta_{pq}\tau_{ij}+\delta_{iq}\tau_{jp}+\delta_{jp}\tau_{iq})\notag
\\[1ex]
&+&
\beta_2(\delta_{ip}\tau_{jq}+\delta_{jq}\tau_{ip})
+
\beta_3\tau_{ip}\tau_{jq}
+
\gamma_1(\delta_{ij}\tau_{pk}\tau_{kq} + \delta_{pq}\tau_{ik}\tau_{kj} + \delta_{iq}\tau_{jk}\tau_{kp} + \delta_{jp}\tau_{ik}\tau_{kq})\notag
\\[1ex]
&+&
\gamma_2(\delta_{ip}\tau_{jk}\tau_{kq} + \delta_{jq}\tau_{ik}\tau_{kp})
+
\gamma_3(\tau_{ip}\tau_{jk}\tau_{kq}+\tau_{jq}\tau_{ik}\tau_{kp})
+
\gamma_4\tau_{ik}\tau_{kp}\tau_{jk}\tau_{kq},
\label{A0-incomp-components-ref}
\end{eqnarray}
since the term involving $\alpha_2$ may be dropped because it vanishes from the expression for the incremental stress by virtue of the incompressibility condition.
Here $\alpha_1$, $\beta_1$, $\beta_3$, $\gamma_1$, $\gamma_3$ and $\gamma_4$ are unchanged, while $\beta_2$ reduces to $4(W_{16} + 2W_{17} + 2W_{26} + 4W_{27})$ and $\gamma_2$ reduces to $4(W_{18} + 2W_{19} + 2W_{28} + 4W_{29})$.
The terms $\beta_2\delta_{jq}\tau_{ip}$ and $\gamma_2\delta_{jq}\tau_{ik}\tau_{kp}$ also vanish from the incremental stress for the same reason but are retained here since otherwise the major symmetry would be lost.
Note that the terms involving $\delta_{jp}$ in \eqref{A0-incomp-components-ref} do not contribute to the equation of motion, again by virtue of the incompressibility condition.

%%%%%%%%%%%%%%%%%%%%%%%%%%%%%

\section{The effect of initial stress on infinitesimal wave propagation\label{sec5}}

%%%%%%%%%%%%%%%%%%%%%%%%%%%%%

Now consider incremental motions in an infinite homogeneous medium subject to homogeneous deformation and/or homogeneous initial stress.  From Section \ref{sec3} we recall that the equation of incremental motion is
\begin{equation}
\mathcal{A}_{0piqj}u_{j,pq}=\rho u_{i,tt},
 \label{5motion-comp}
\end{equation}
for a compressible material, and
\begin{equation}
\mathcal{A}_{0piqj}u_{j,pq}-\dot{p}_{,i}=\rho u_{i,tt},\quad u_{i,i}=0,
\label{5motion-incomp}
\end{equation}
for an incompressible material.

Consider a homogeneous plane wave of the form
\begin{equation}
\mathbf{u}=\mathbf{m}f(\mathbf{n}\cdot\mathbf{x}- v t),
\end{equation}
where $\mathbf{m}$ is a fixed unit vector (the polarization vector), $f$ is an arbitrary function of the argument $\mathbf{n}\cdot\mathbf{x}- v t$, $\mathbf{n}$ is a unit vector in the direction of propagation, and $v$ is the wave speed.
For an incompressible material we also set
\begin{equation}
\dot{p}=g(\mathbf{n}\cdot\mathbf{x} - v t),
\end{equation}
where $g$ is a function related to $f$. Substitution into the equation of motion \eqref{5motion-comp} (after dropping $f''$, which is assumed to be non-zero) leads to
\begin{equation}
\mathcal{A}_{0piqj}n_pn_qm_j=\rho v^2 m_i,\label{prop-cond-compx}
\end{equation}
for a compressible material, and into \eqref{5motion-incomp}
\begin{equation}
\mathcal{A}_{0piqj}n_pn_qm_j f''-g'n_i =\rho v^2 m_i f'',\qquad m_in_i=0,\label{plane-incomp}
\end{equation}
for an incompressible material, where a prime on $f$ or $g$ indicates the derivative with respect to its argument.
By multiplying the first equation in \eqref{plane-incomp} by $n_i$ and using the second equation, we obtain
\begin{equation}
g'=\mathcal{A}_{0piqj}n_pn_qm_jn_i f'',
\end{equation}
and upon substitution back in \eqref{plane-incomp} for $g'$ and again dropping $f''$ we arrive at
\begin{equation}
(\mathcal{A}_{0piqj}n_pn_q-\mathcal{A}_{0pkqj}n_pn_qn_kn_i)m_j=\rho v^2 m_i.\label{prop-cond-incompx}
\end{equation}

It is convenient now to use the \emph{acoustic tensor} $\mathbf{Q}(\mathbf{n})$, whose components are defined by
\begin{equation}
Q_{ij}(\mathbf{n})=\mathcal{A}_{0piqj}n_pn_q.\label{acoustic-tensor}
\end{equation}
Then equations \eqref{prop-cond-compx} and \eqref{prop-cond-incompx} can be written in the compact forms
\begin{equation}
\mathbf{Q}(\mathbf{n})\mathbf{m}=\rho v^2\mathbf{m},\label{prop-cond-comp}
\end{equation}
\begin{equation}
\mathbf{\bar{Q}}(\mathbf{n})\mathbf{m}=\rho v^2\mathbf{m},\qquad \mathbf{m}\cdot\mathbf{n}=0,\label{prop-cond-incomp}
\end{equation}
for unconstrained and incompressible materials, respectively, where, noting that $\mathbf{\bar{I}}\mathbf{m}=\mathbf{m}$, with $\mathbf{\bar{I}}=\mathbf{I}-\mathbf{n}\otimes\mathbf{n}$ defined as the projection operator on to the plane normal to $\mathbf{n}$, the symmetric tensor $\mathbf{\bar{Q}}(\mathbf{n})$ is defined as
\begin{equation}
\mathbf{\bar{Q}}(\mathbf{n})=\mathbf{\bar{I}}\mathbf{Q}(\mathbf{n})\mathbf{\bar{I}}.\label{Q-project}
\end{equation}
This is the projection of $\mathbf{Q}(\mathbf{n})$ on to the plane normal to $\mathbf{n}$.  The symmetrization of the acoustic tensor in the incompressible case is attributed to Hayes in the Ph.D. thesis of \cite{Scot74}\footnote{We are grateful to a reviewer for bringing this reference to our attention.}; see also \cite{Scot85} and \cite{Boul93}.

For any given direction of propagation $\mathbf{n}$ we now have a symmetric algebraic eigenvalue problem for determining $\rho v^2$ and $\mathbf{m}$, which is three-dimensional in the compressible case and two-dimensional in the incompressible case.
Because of the symmetry there are three mutually orthogonal eigenvectors $\mathbf{m}$ corresponding to the direction of displacement in the compressible case and two in the incompressible case, corresponding to transverse waves.  The values of $\rho v^2$, three and two, respectively, are obtained from the characteristic equations
\begin{equation}
\det[\mathbf{Q}(\mathbf{n})-\rho v^2 \mathbf{I}]=0,\qquad \det[\mathbf{\bar{Q}}(\mathbf{n})-\rho v^2 \mathbf{\bar{I}}]=0,\label{secular}
\end{equation}
for compressible and incompressible materials, respectively.

If $\mathbf{m}$ is known then $\rho v^2$ is given by
\begin{equation}
\rho v^2=[\mathbf{Q}(\mathbf{n})\mathbf{m}]\cdot\mathbf{m},
\end{equation}
which applies for either compressible or incompressible materials.

To examine plane waves in detail we need to calculate $\mathbf{Q}(\mathbf{n})$ for the forms of $\boldsymbol{\mathcal{A}}_0$ given in Section \ref{sec4-3}.
We now do this explicitly for the reference configuration $\mathcal{B}_r$.
From \eqref{A0-comp-components-ref} and \eqref{acoustic-tensor} we obtain
\begin{eqnarray}
\mathbf{Q}(\mathbf{n}) &=&
\left[\alpha_1 + (1+\beta_1)(\mathbf{n\cdot \bm{\tau} n}) + \gamma_1(\mathbf{n\cdot \bm\tau^2 n})\right]\mathbf{I}+\beta_1 \boldsymbol{\tau}+\gamma_1 \boldsymbol{\tau}^2\notag
\\[0.1cm]
&+&(\alpha_1+\alpha_2)\mathbf{n}\otimes\mathbf{n}+
(\beta_1+\beta_2) (\mathbf{n}\otimes\boldsymbol{\tau n}+\boldsymbol{\tau n}\otimes\mathbf{n}) +\beta_3 \boldsymbol{\tau n}\otimes\boldsymbol{\tau n}\notag
\\[0.1cm]
&+&(\gamma_1+ \gamma_2) (\mathbf{n}\otimes\boldsymbol{\tau}^2 \mathbf{n} + \boldsymbol{\tau}^2 \mathbf{n}\otimes\mathbf{n})+\gamma_3 (\bm\tau \mathbf{n}\otimes\boldsymbol{\tau}^2 \mathbf{n} + \boldsymbol{\tau}^2 \mathbf{n}\otimes\bm\tau\mathbf{n})+
\gamma_4 \boldsymbol{\tau}^2\mathbf{n}\otimes \boldsymbol{\tau}^2 \mathbf{n}\notag\\
 \label{Q-comp-ref}
\end{eqnarray}
for compressible materials, and from \eqref{A0-incomp-components-ref}, \eqref{acoustic-tensor}, and \eqref{Q-project}, we obtain
\begin{eqnarray}
\mathbf{\bar{Q}}(\mathbf{n})
&=&
\left[\alpha_1 + (1+\beta_1)(\mathbf{n\cdot \bm{\tau} n}) + \gamma_1(\mathbf{n\cdot \bm\tau^2 n})\right]\mathbf{\bar{I}} + \beta_1 \mathbf{\bar{I}}\bm\tau \mathbf{\bar{I}}
+ \gamma_1 \mathbf{\bar{I}}\bm\tau^2 \mathbf{\bar{I}}\notag
\\[0.1cm]
&+&\beta_3 \mathbf{\bar{I}}\boldsymbol{\tau n}\otimes\mathbf{\bar{I}}\boldsymbol{\tau n}
+\gamma_3 \left[\mathbf{\bar{I}}\boldsymbol{\tau n}\otimes\mathbf{\bar{I}}\boldsymbol{\tau}^2 \mathbf{n} + \mathbf{\bar{I}}\boldsymbol{\tau}^2 \mathbf{n}\otimes\mathbf{\bar{I}}\boldsymbol{\tau} \mathbf{n}\right]
+ \gamma_4 \mathbf{\bar{I}}\boldsymbol{\tau}^2 \mathbf{n}\otimes\mathbf{\bar{I}}\boldsymbol{\tau}^2 \mathbf{n}
\label{Q-incomp-ref}
\end{eqnarray}
for incompressible materials.
Note, in particular, the explicit nonlinear dependence of the acoustic tensors  on the initial stress $\bm\tau$ (not forgetting that the $\alpha$'s, $\beta$'s, and $\gamma$'s may themselves be nonlinear in $\bm\tau$).

%------------------------------

\paragraph{Example 1:  Longitudinal and transverse waves in compressible materials}

%------------------------------

We consider the possibility of a pure longitudinal wave in the reference configuration (initial stress, no pre-strain), so that $\mathbf{m}=\mathbf{n}$.
From \eqref{prop-cond-comp} and \eqref{Q-comp-ref} we obtain
\begin{equation}
(a - \rho v^2)\mathbf{n} + b\bm\tau\mathbf{n} + c\bm\tau^2\mathbf{n} = \mathbf{0},\label{long-prop-cond}
\end{equation}
where the constants $a$, $b$, $c$ are given by
\begin{align}
& a = 2\alpha_1 + \alpha_2 + (1 + 2\beta_1 + \beta_2)(\mathbf{n}\cdot \bm\tau\mathbf{n})+ (\gamma_1 + \gamma_2)(\mathbf{n}\cdot \bm\tau^2\mathbf{n}),
\notag \\[0.1cm]
& b = 2\beta_1 + \beta_2 + \beta_3 (\mathbf{n}\cdot\bm\tau\mathbf{n}) + \gamma_3(\mathbf{n}\cdot \bm\tau^2\mathbf{n}),
\notag \\[0.1cm]
& c=   2\gamma_1 + \gamma_2 + \gamma_3 (\mathbf{n}\cdot\bm\tau\mathbf{n}) + \gamma_4 (\mathbf{n}\cdot \bm\tau^2\mathbf{n}).
\end{align}
Several distinct cases arise for which a longitudinal wave can propagate provided the material parameters and the initial stresses are such that $\rho v^2>0$.  First, for the degenerate case in which $b=c=0$ we have $\rho v^2=a$ and a longitudinal wave can propagate in any direction.  Second, if $c=0$ and $b\neq 0$ then $\mathbf{n}$ is an eigenvector of $\bm\tau$.  Recalling that $\tau_i,\,i=1,2,3$, are the principal values of $\bm\tau$, a longitudinal wave can then propagate along the $i$-th principal direction, $i=1,2,3$, with speed $v$ given by $\rho v^2=a+b\tau_i$.  Third, if $c\neq 0$ then we refer \eqref{long-prop-cond} to the principal axes of $\bm\tau$, leading to
\begin{equation}
(a-\rho v^2+b\tau_i+c\tau_i^2)n_i=0,\quad i=1,2,3.
\end{equation}
There are now different possibilities depending on the multiplicity of the eigenvalues of $\bm\tau$.  If $\tau_i=\tau$ say, $i=1,2,3$, then $\rho v^2=a+b\tau+c\tau^2$ and a longitudinal wave can propagate in any direction.  If two of the eigenvalues are equal but distinct from the third, say $\tau_1=\tau_2=\tau\neq\tau_3$, then a longitudinal wave can propagate along the 3 direction with speed given by $\rho v^2=a+b\tau_3+c\tau_3^2$ or in the principal $(1,2)$ plane with speed given by $\rho v^2=a+b\tau+c\tau^2$.  There is also a special case in which if $b+c(\tau+\tau_3)=0$ then a longitudinal wave can propagate in an arbitrary direction with speed $\rho v^2=a-c\tau\tau_3$.  Finally, if the eigenvalues $\tau_i$ are distinct then, for propagation in the $i$-th principal direction, the wave speed, expanded in full, is given by
\begin{equation} \label{longi}
\rho v^2 = 2\alpha_1 + \alpha_2 + (1 + 4\beta_1 + 2\beta_2)\tau_i + (\beta_3 + 3\gamma_1 + 2\gamma_2) \tau_i^2 + 2\gamma_3\tau_i^3 + \gamma_4 \tau_i^4.
\end{equation}
It is also easy to show that propagation is possible in any direction within an $(i,j)$ principal plane if $b+c(\tau_i+\tau_j)=0$, in which case the formula for the wave speed is $\rho v^2=a-c\tau_i\tau_j$.

Of course, a longitudinal wave is in general accompanied by a pair of transverse waves with mutually orthogonal polarizations.
In the case where a pure longitudinal wave exists and $\mathbf{n}$ is aligned with the principal axis $\mathbf{e}_i$ of $\bm\tau$, say, with $i=1,2$ or $3$, it follows that the expression \eqref{Q-comp-ref} for the acoustical tensor reduces to 
\begin{eqnarray}
\mathbf{Q}(\mathbf{e}_i)&=&
[\alpha_1+(1+\beta_1)\tau_{i} + \gamma_1 \tau_i^2]\mathbf{I}
+\beta_1\boldsymbol{\tau}  + \gamma_1\bm\tau^2\notag
\\[0.1cm]
&+&
\left[\alpha_1 + \alpha_2 + 2(\beta_1 + \beta_2)\tau_{i} + (\beta_3 + 2\gamma_1 + 2\gamma_2) \tau_{i}^2 + 2\gamma_3\tau_i^3 + \gamma_4\tau_i^4 \right] \mathbf{e}_i\otimes\mathbf{e}_i .\qquad
\end{eqnarray}
Hence, for a corresponding transverse wave with polarization $\mathbf{m}$ and speed $v$, we have, on use of \eqref{prop-cond-comp},
\begin{equation}
[\alpha_1+(1+\beta_1)\tau_{i} + \gamma_1 \tau_i^2]\mathbf{m}+\beta_1\bm\tau\mathbf{m} + \gamma_1\bm\tau^2\mathbf{m} =\rho v^2\mathbf{m}.\label{transexist-prev}
\end{equation}
In general it does not follow that $\mathbf{m}$ is also a principal axis of $\bm\tau$. However, in the special case in which $\mathbf{m}$ \emph{is} also a principal axis of $\bm\tau$ (in the plane normal to $\mathbf{e}_i$), let this correspond to principal initial stress $\tau_j,\,j\neq i$. Then, the wave speed, denoted $v_{ij}$, associated with $\mathbf{m}=\mathbf{e}_j$ is given by
\begin{equation}
\rho v_{ij}^2=\tau_i+\alpha_1+\beta_1(\tau_{i} +\tau_j) + \gamma_1 (\tau_i^2 +\tau_j^2).
\end{equation}
The difference
\begin{equation}
\rho (v_{ij}^2 - v_{ji}^2) = \tau_i -\tau_j ,
\end{equation}
{\color{black}recalls the universal relationship established by \cite{Man87} for the general linear theory.}  Note, however, that in \cite{Man87} the material was taken to be orthotropic with the principal axes of $\bm\tau$ coinciding with the axes of orthotropy, whereas here the orthotropy is associated with $\bm\tau$ itself and the underlying material is isotropic. The principal axes are eigenvectors of $\mathbf{Q}$ in each case although the values of the components $Q_{ij}$ differ.

The situation is somewhat different if one starts by considering the existence of a pure transverse wave.
There is then no general result showing that such a wave must be travelling and/or polarized along a principal axis of initial stress.
We do not include here the details of this case, and refer instead to \cite{Man87}  for an example of a wave polarized along one principal direction of stress and propagating in any direction in the plane normal to the polarization.

In all the above examples, for a real wave to exist it is necessary that $\rho v^2>0$, which puts restrictions on the parameters and initial stress components involved.  In general this requirement may be stated in terms of the  \emph{strong ellipticity condition}, i.e.
\begin{equation}
\mathbf{m}\cdot[\mathbf{Q}(\mathbf{n})\mathbf{m}]>0\quad \mbox{for all non-zero vectors}\ \mathbf{m}, \mathbf{n}.
\end{equation}
This also applies in the incompressible case, subject to the restriction $\mathbf{m}\cdot\mathbf{n}=0$.

%-------------------------

\paragraph{Example 2:  {\color{black} Connections with} the results of Man and Lu}

%-------------------------

Suppose now that in \eqref{Q-comp-ref} the coefficients $\alpha_1$, $\alpha_2$, $\beta_1$, $\beta_2$ are constants and only the terms that are linear in the initial stress $\bm\tau$ are retained. In particular, the terms with coefficients $\beta_3$, $\gamma_i$ ($i=1,2,3,4$) are omitted.  Then, referred to the principal axes of $\bm\tau$, the components of $\mathbf{Q}(\mathbf{n})$ can be written compactly as
\begin{align}
& Q_{ii} = \alpha_1+(\alpha_1+\alpha_2)n_i^2+(1+\beta_1)(\mathbf{n} \cdot \bm\tau\mathbf{n})+\beta_1\tau_i+2(\beta_1+\beta_2)\tau_in_i^2,\quad i=1,2,3,
\notag \\[0.1cm]
& Q_{ij} = [\alpha_1+\alpha_2+(\beta_1+\beta_2)(\tau_i+\tau_j)]n_in_j,\quad i\neq j,
\end{align}
where $\tau_i$ ($i=1,2,3$) are the principal values of $\bm\tau$.
These formulas {\color{black} are consistent} with corresponding expressions for the components of the acoustic tensor (or Christoffel tensor) obtained by \cite{Man87} [their eq. (18)] although this is not immediately clear since the notation differs.
The correspondence can be established by noting that the tensor $\bm{\mathcal{L}}$ used by \cite{Man87} has components that are connected to those of $\bm{\mathcal{A}}_0$ by $\mathcal{L}_{ijkl}=\mathcal{A}_{0ijkl}-\delta_{jl}\tau_{ik}$.

%-------------------------

\paragraph{Example 3:  {\color{black} Connections with} the classical theory of Biot}

%-------------------------

In the classical theory of \citet{Biot39,Biot40,Biot65} the incremental stress may be written (in the present notation) as $\bm{\mathcal{B}}\mathbf{L}$ or in index notation as $\mathcal{B}_{piqj}u_{j,q}$, where the coefficients $\mathcal{B}_{piqj}$ satisfy the relations
\begin{equation}
\mathcal{B}_{piqj}=\mathcal{B}_{ipqj}=\mathcal{B}_{pijq},\quad \mathcal{B}_{piqj}-\mathcal{B}_{qjpi}=\delta_{ip}\tau_{jq}-\delta_{jq}\tau_{ip},\label{Biot-conds}
\end{equation}
the latter following from the existence of a strain-energy function.
It was shown in \cite{Ogde11} that the general connection between $\mathcal{A}_{0piqj}$ and $\mathcal{B}_{piqj}$ may be written in the form
\begin{equation}
\mathcal{A}_{0piqj} = \mathcal{B}_{piqj} - \tfrac{1}{2} \delta_{pj} \tau_{qi} - \tfrac{1}{2} \delta_{pq} \tau_{ij} - \tfrac{1}{2} \delta_{qi} \tau_{pj} + \tfrac{1}{2}  \delta_{ij} \tau_{pq} + \delta_{qj} \tau_{pi}.
\end{equation}
It is worth noting that this connection does not depend on the existence of a strain-energy function and is independent of any material symmetry.
For the general expression \eqref{A0-comp-components-ref} to reduce to the Biot form for isotropic response the material parameters in \eqref{A0-comp-components-ref} must be specialized to
\begin{equation}
\alpha_1=\mu, \qquad
\alpha_2=\lambda, \qquad
\beta_1=-1/2, \qquad
\beta_2=1, \qquad
\label{biot1}
\end{equation}
where $\mu$ and $\lambda$ are the Lam\'e moduli, and the terms which are of order higher than 1 in $\bm\tau$ must be neglected.
This corresponds to $\mathcal{B}_{piqj}$ having the form
\begin{equation}
\mathcal{B}_{piqj}=\mu(\delta_{ij}\delta_{pq}+\delta_{qi}\delta_{pj})+\lambda\delta_{pi}\delta_{qj}+\delta_{pi}\tau_{qj},\label{biot-iso}
\end{equation}
which satisfies the conditions \eqref{Biot-conds}.

Clearly, {\color{black} the formula \eqref{biot-iso}} relies on the special identifications \eqref{biot1}, and is very much a specialization of the more general theory discussed here.
In effect, if one was to linearize the general expression of the elastic moduli  \eqref{A0-comp-components-ref} with respect to the initial stress, then the expansions
\begin{equation}
\alpha_1= \mu + \hat{\alpha}_1 \tau, \qquad
\alpha_2=\lambda + \hat{\alpha}_2 \tau, \qquad
\beta_1= \hat{\beta}_1, \qquad
\beta_2= \hat{\beta_2},
\end{equation}
would be required,
where ``$\tau$'' is a term of first order in $\bm \tau$, and the scalars with a hat are constants, which cannot be determined a priori and must be measured experimentally.
In conclusion, six material constants ($\lambda$, $\mu$,  $\hat{\alpha}_1$, $\hat{\alpha}_2$, $\hat{\beta}_1$, $\hat{\beta}_2$) are required to describe the isotropic response of a compressible material with a small initial stress, not just the two Lam\'e coefficients as Biot implied.
We refer to \cite{Man98} for a discussion of this point in relation to Hartig's law and a proof that four additional constants are needed to describe the elastic response of currently known real materials.

%-------------------------

\paragraph{Example 4:  Second-order elasticity}

%-------------------------

Here we consider an isotropic elastic solid without initial stress and its specialization to second order in the components of the Green strain tensor $\mathbf{E}=\frac{1}{2}(\mathbf{C}-\mathbf{I})$, where $\mathbf{C}$ is again the right Cauchy--Green deformation tensor.  An appropriate form of the elastic strain-energy function, which is correct to second order, is that due to \cite{Murn}.  When expressed in terms of the invariants of $\mathbf{C}$ given by \eqref{invs1} the Murnaghan energy function has the form
\begin{eqnarray}
W &=&
\dfrac{\lambda}{8} (I_1 - 3)^2 + \dfrac{\mu}{4}(I_1^2 - 2I_1 - 2I_2 + 3)\notag
\\[1ex]
&+& \dfrac{l}{24}(I_1 - 3)^3
+ \dfrac{m}{12}(I_1 - 3)(I_1^2 - 3I_2) + \dfrac{n}{8}(I_1 - I_2 + I_3 - 1),
\label{Murnaghan}
\end{eqnarray}
where $\lambda$ and $\mu$ are the classical Lam\'e constants of linear elasticity and $l$, $m$, $n$ are the second-order constants of Murnaghan.  For this energy function we have $W_{22}=W_{13}=W_{23}=W_{33}=0$ and since we are not including initial stress the expression \eqref{mathcalA0-components-comp} reduces to
\begin{eqnarray}
J\mathcal{A}_{0piqj}
&=&2(W_1+I_1W_2)B_{pq}\delta_{ij}+2W_2[2B_{pi}B_{qj}-B_{iq}B_{jp}-B_{pr}B_{rq}\delta_{ij}-B_{pq}B_{ij}]
\notag\\[0.1cm]
 &+& 2I_3W_3(2\delta_{ip}\delta_{jq}-\delta_{iq}\delta_{jp})+4W_{11}B_{ip}B_{jq}\notag\\[0.1cm]
&+& 4W_{12}(2I_1B_{ip}B_{jq}-B_{ip}B_{jr}B_{rq}-B_{jq}B_{ir}B_{rp}),
\end{eqnarray}
and the remaining coefficients $W_1$, $W_2$, $W_3$, $W_{11}$ and $W_{12}$ are simply obtained from \eqref{Murnaghan}.  Then, with this specialization, the wave speed $v$ is given by
\begin{equation}
\rho v^2= \mathcal{A}_{0piqj}n_pn_qm_im_i.
\end{equation}
In working with second-order elasticity the corrections to the classical longitudinal and transverse wave speeds are obtained to first order in $\mathbf{E}$.  We have $\mathbf{C}=\mathbf{I}+2\mathbf{E}$ and $I_1=3+2E$, exactly, which we use together with the linear approximations $I_2 \simeq 3+4E$, $I_3 \simeq 1 +2E$, $\mathbf{B} \simeq \mathbf{C}=\mathbf{I}+2\mathbf{E}$, and we also note that $\rho \simeq \rho_r(1-E)$, where $E=\tr\mathbf{E}$.

To the first order in $\mathbf{E}$ we then obtain
\begin{eqnarray}
W_1&=&
\mu+\tfrac{1}{8}n+\tfrac{1}{2}(\lambda+2\mu+2m)E,\quad W_2=-\tfrac{1}{2}\mu-\tfrac{1}{8}n-\tfrac{1}{2}mE,\quad W_3=\tfrac{1}{8}n,
\notag \\[0.1cm]
W_{11}
&=& \tfrac{1}{4}(\lambda+2\mu+4m)+\tfrac{1}{2}(l+2m)E,\quad W_{12}=-\tfrac{1}{4}m.
\end{eqnarray}

After some manipulations the above approximations enable us to obtain, to the first order in $\mathbf{E}$,
\begin{eqnarray}
J\mathcal{A}_{0piqj}
&=&
\mu(\delta_{ij}\delta_{pq}+\delta_{iq}\delta_{jp})+\lambda\delta_{ip}\delta_{jq}+2\mu (2\delta_{ij}E_{pq}+\delta_{pq}E_{ij}+\delta_{iq}E_{jp}+\delta_{jp}E_{iq})\notag\\[0.1cm]
&+&\lambda(E\delta_{ij}\delta_{pq}+2\delta_{ip}E_{jq}+2\delta_{jq}E_{ip})+2lE\delta_{ip}\delta_{jq}\notag\\[0.1cm]
&+& m[E(\delta_{ij}\delta_{pq}+\delta_{iq}\delta_{jp}-2\delta_{ip}\delta_{jq})+2(\delta_{ip}E_{jq}+\delta_{jq}E_{ip})]\notag\\[0.1cm]
&+&\tfrac{1}{2}n[\delta_{ij}E_{pq}+\delta_{pq}E_{ij}+\delta_{iq}E_{jp}+\delta_{jp}E_{iq}-2\delta_{ip}E_{jq}-2\delta_{jq}E_{ip}\notag\\[0.1cm]
&-&E(\delta_{ij}\delta_{pq}+\delta_{iq}\delta_{jp}-2\delta_{ip}\delta_{jq})],
\end{eqnarray}
and hence, for any $\mathbf{m}$, $\mathbf{n}$ pair satisfying \eqref{prop-cond-comp}, the wave speed is obtained via
\begin{eqnarray}
\rho_r v^2
&=&
\mu +
(\mu+\lambda)(\mathbf{m}\cdot\mathbf{n})^2
+ 2\mu [2(\mathbf{n\cdot En}) + (\mathbf{m}\cdot\mathbf{Em}) + 2(\mathbf{m}\cdot\mathbf{n})
(\mathbf{m}\cdot\mathbf{En})]
\notag\\[0.1cm]
&
+&\lambda [E + 4(\mathbf{m}\cdot\mathbf{n})
(\mathbf{m}\cdot\mathbf{En})]
+ 2lE(\mathbf{m}\cdot\mathbf{n})^2
\notag\\[0.1cm]
&
+& m\{E[1-(\mathbf{m}\cdot\mathbf{n})^2] + 4(\mathbf{m}\cdot\mathbf{n})(\mathbf{m}\cdot\mathbf{En}) \}
\notag\\[0.1cm]
&
+&\tfrac{1}{2}n\{(\mathbf{n}\cdot\mathbf{En}) + (\mathbf{m}\cdot\mathbf{Em}) - 2(\mathbf{m}\cdot\mathbf{n})
(\mathbf{m}\cdot\mathbf{En}) -E[1-(\mathbf{m}\cdot\mathbf{n})^2]\}.
\end{eqnarray}
For a longitudinal wave with $\mathbf{m}=\mathbf{n}$ this reduces to
\begin{equation}
\rho_r v^2=\lambda+2\mu+(\lambda+2l)E+(10\mu+4\lambda+4m)(\mathbf{n}\cdot\mathbf{En}),
\end{equation}
and for a transverse wave with $\mathbf{m}\cdot\mathbf{n}=0$ and $\mathbf{m}\times\mathbf{n}=\mathbf{l}$,
\begin{equation}
\rho_r v^2=\mu+2\mu (\mathbf{n}\cdot\mathbf{En}) + (\lambda+2\mu+m)E - \tfrac{1}{2}(4\mu+n)(\mathbf{l}\cdot\mathbf{El}),
\end{equation}
where we have made use of the connection $\mathbf{l}\cdot\mathbf{El} + \mathbf{m}\cdot\mathbf{Em} + \mathbf{n}\cdot\mathbf{En} = E$.
When $\mathbf{n}$ is specialized to the axis $\mathbf{e}_1$ and $\mathbf{m}$ is taken to be $\mathbf{e}_2$ and $\mathbf{e}_3$ in turn, where $\mathbf{e}_1$, $\mathbf{e}_2$ and $\mathbf{e}_3$ are principal axes of strain, then these results agree with those obtained by \cite{HuKe53} [their eq. (11)].

It is straightforward to show that a longitudinal wave actually exists if \emph{either} $\mathbf{n}$ is an eigenvector of $\mathbf{E}$ \emph{or} the elastic constants satisfy $\lambda+2\mu+m=0$.
For the existence of a transverse wave there are more options.
First, in the special case when the elastic constants are such that $\lambda+2\mu+m=0$ \emph{and}
$4\mu+n=0$, a transverse wave exists for any direction of propagation $\mathbf{n}$.  Second, if $4(\lambda+\mu+m)-n=0$ and $4\mu+n\neq 0$ then a transverse wave exists only if $\mathbf{m}$ is an eigenvector of $\mathbf{E}$.
Third, if $4\mu+n=0$ and $\lambda+2\mu+m\neq 0$ then a transverse wave exists for any $\mathbf{n}$ for which $\mathbf{Em}$ lies in the plane normal to $\mathbf{n}$.  Finally, if none of these very special cases apply then the existence of a transverse wave requires that $\mathbf{m}$ be an eigenvector of $\mathbf{E}$.

Note that the case of incompressible materials is treated elsewhere; see \cite{Destrade10} and references therein.

%------------------

\paragraph{Example 5:  A simple nonlinear model}

%------------------

The above examples have involved different specializations that consider the strain and/or the initial stress to be small.
In this final example we consider a finite deformation from a configuration that is subject to an initial stress that is nonlinear.
For this purpose it suffices to adopt a simple prototype incompressible form of strain-energy function, which is taken to be
\begin{equation}
W=\tfrac{1}{2}\mu (I_1-3)+\tfrac{1}{2}\bar{\mu}(I_6-\tr\bm\tau)^2+\tfrac{1}{2}(I_6-\tr\bm\tau),
\end{equation}
where the first term represents the classical (isotropic) strain-energy function of a neo-Hookean material with shear modulus $\mu$, while the terms in $I_6$ capture the effect of initial stress and satisfy \eqref{ref-conds-incomp}$_2$.  The constant $\bar{\mu}$ has dimensions of [stress]$^{-1}$.  We recall that $I_6=\tr(\mathbf{C}\bm\tau)$, which may also be written as $\tr\bm\Sigma$.

From \eqref{incomp-sigma} the Cauchy stress is then obtained as
\begin{equation}
\bm\sigma=\mu\mathbf{B}-p\mathbf{I}+[1+2\bar{\mu}(I_6-\tr\bm\tau)]\bm\Sigma,\label{Cauchy-proto}
\end{equation}
and from \eqref{mathcalA0-components-incomp} the components of $\bm{\mathcal{A}}_0$ reduce to
\begin{equation}
\mathcal{A}_{0piqj}=\mu B_{pq}\delta_{ij}+[1+2\bar{\mu}(I_6-\tr\bm\tau)]\Sigma_{pq}\delta_{ij}+4\bar{\mu}\Sigma_{pi}\Sigma_{qj}.
\end{equation}
The corresponding components of the acoustic tensor are obtained from \eqref{acoustic-tensor}, and it is convenient to write it in tensor notation as
\begin{equation}
\mathbf{Q}(\mathbf{n})=\alpha \mathbf{I}+\beta \bm\Sigma\mathbf{n}\otimes\bm\Sigma\mathbf{n},
\end{equation}
where for compactness of representation we have introduced the notations
\begin{equation}
\alpha=\mu(\mathbf{n}\cdot\mathbf{Bn}) + [1+2\bar{\mu}(I_6-\tr\bm\tau)](\mathbf{n}\cdot\mathbf{\bm\Sigma n}),
\qquad \beta =4\bar{\mu}.
\end{equation}

Since we are considering an incompressible material we form the projection of $\mathbf{Q}(\mathbf{n})$ on to the plane normal to $\mathbf{n}$ according to \eqref{Q-project}.  This is
\begin{equation}
\mathbf{\bar{Q}}(\mathbf{n})=\alpha\mathbf{\bar{I}}+\beta\mathbf{\bar{I}}\bm\Sigma\mathbf{n}\otimes\mathbf{\bar{I}}\bm\Sigma\mathbf{n}.
\end{equation}
Next we solve \eqref{secular}$_2$ and find that the two solutions for $\rho v^2$ are
\begin{equation}
\rho v^2=\alpha,\qquad \rho v^2=\alpha +\beta (\mathbf{n}\cdot\mathbf{\bm\Sigma m})^2,
\end{equation}
and equation \eqref{prop-cond-incomp} becomes
\begin{equation}
\alpha\mathbf{m}
+
\beta (\mathbf{n}\cdot\mathbf{\bm\Sigma m}) \mathbf{\bar{I}}\bm\Sigma\mathbf{n}=\rho v^2\mathbf{m},\qquad \mathbf{m}\cdot\mathbf{n}=0.
\end{equation}
Assuming that $\beta\neq0$, for the solution $\rho v^2=\alpha$ this requires that either $\mathbf{\bar{I}}\bm\Sigma\mathbf{n}=\mathbf{0}$ or $\mathbf{n}\cdot\mathbf{\bm\Sigma m}=0$.
If $\mathbf{\bar{I}}\bm\Sigma\mathbf{n}=\mathbf{0}$ then $\mathbf{n}$ is an eigenvector of $\bm\Sigma$ and both transverse waves have the same speed $v$, given by $\rho v^2=\alpha$.
If, on the other hand, $\mathbf{n}\cdot\mathbf{\bm\Sigma m}=0$ and $\mathbf{\bar{I}}\bm\Sigma\mathbf{n}\neq\mathbf{0}$ then the wave speed corresponding to polarization $\mathbf{m}$ is given by $\rho v^2=\alpha$, while for the second transverse wave, with polarization $\mathbf{m}'$ say, the wave speed is given by $\rho v^2=\alpha +\beta(\mathbf{n}\cdot\mathbf{\bm\Sigma m}')^2$.

As a simple illustration of the above results we now consider the initial stress to be uniaxial, of the form $\bm\tau=\tau_1\mathbf{e}_1\otimes\mathbf{e}_1$, where $\mathbf{e}_1$ is a principal axis of $\mathbf{B}$ corresponding to principal stretch $\lambda_1$. Then, $\bm\Sigma=\tau_1\lambda_1^2\mathbf{e}_1\otimes\mathbf{e}_1$.  It follows that $\mathbf{n}\cdot\mathbf{\bm\Sigma m}=0$ and
\begin{equation}
\rho v^2=(\mu+\tau_1)\lambda_1^2+2\bar{\mu}\tau_1^2\lambda_1^2(\lambda_1^2-1).
\end{equation}
It now suffices to measure the wave speed for three different values of $\lambda_1$ to evaluate the material constants $\mu$ and $\bar\mu$, and the initial stress $\tau_1$.
If the deformation is achieved by applying a uniaxial stress along $\mathbf{e}_1$ then the associated principal Cauchy stress $\sigma_1$ is obtained from \eqref{Cauchy-proto} as
\begin{equation}
\sigma_1=(\mu+\tau_1)\lambda_1^2-p+2\bar{\mu}\tau_1^2\lambda_1^2(\lambda_1^2-1).
\end{equation}
By symmetry, incompressibility and vanishing of the lateral stress it also follows from \eqref{Cauchy-proto} that $p=\mu\lambda_1^{-1}$.  If this is used in the above then the expression for the wave speed can be written simply as $\rho v^2=\sigma_1+\mu\lambda_1^{-1}$.

%%%%%%%%%%%%%

\appendix

\section*{Appendix A}

%+++++++++++++++++++++++++++++++++++++++++++++++++++

\subsection*{A1. First derivatives of the invariants}

%+++++++++++++++++++++++++++++++++++++++++++++++++++

Here we list the first derivatives of the invariants $I_1,I_2,I_3,I_6,\dots, I_9$ with respect to the deformation gradient.
These are
\begin{align}
&
\dfrac{\partial I_1}{\partial \mathbf{F}} = 2 \mathbf{F}^\mathrm{T},
&&
\dfrac{\partial I_2}{\partial \mathbf{F}} = 2 (I_1 \mathbf{F}^\mathrm{T} - \mathbf{F}^\mathrm{T}\mathbf{F F}^\mathrm{T}),
&&
\dfrac{\partial I_3}{\partial \mathbf{F}} = 2 I_3 \mathbf{F}^{-1},
\notag \\[4pt]
&
\dfrac{\partial I_6}{\partial \mathbf{F}} = 2 \bm\tau\mathbf{F}^\mathrm{T},
& &
\dfrac{\partial I_7}{\partial \mathbf{F}} = 2  \bm\tau \mathbf{F}^\mathrm{T}\mathbf{F F}^\mathrm{T}
+ 2 \mathbf{F}^\mathrm{T}\mathbf{F}  \bm\tau \mathbf{F}^\mathrm{T},
& &
\notag \\[4pt]
&
\dfrac{\partial I_8}{\partial \mathbf{F}} = 2 \bm\tau^{2}\mathbf{F}^\mathrm{T},
&&
\dfrac{\partial I_9}{\partial \mathbf{F}} = 2  \bm\tau^{2} \mathbf{F}^\mathrm{T}\mathbf{F F}^\mathrm{T}
+ 2 \mathbf{F}^\mathrm{T}\mathbf{F}  \bm\tau^{2} \mathbf{F}^\mathrm{T},
&&
\notag
\end{align}
from which we obtain the expressions
\begin{align}
&
\mathbf{F} \dfrac{\partial I_1}{\partial \mathbf{F}} = 2 \mathbf{B},
&&
\mathbf{F} \dfrac{\partial I_2}{\partial \mathbf{F}} = 2 (I_1 \mathbf{B} - \mathbf{B}^2),
&&
\mathbf{F} \dfrac{\partial I_3}{\partial \mathbf{F}} = 2 I_3 \mathbf{I},
\notag \\[4pt]
& \mathbf{F} \dfrac{\partial I_6}{\partial \mathbf{F}} = 2 \bm\Sigma,
&&
\mathbf{F} \dfrac{\partial I_7}{\partial \mathbf{F}} = 2  \bm\Sigma \mathbf{B}
+ 2 \mathbf{B}  \bm\Sigma,
&&
\notag\\[4pt]
&
\mathbf{F} \dfrac{\partial I_8}{\partial \mathbf{F}} = 2 \bm\Sigma\mathbf{B}^{-1}\bm\Sigma,
&&
\mathbf{F} \dfrac{\partial I_9}{\partial \mathbf{F}} = 2  \bm\Sigma \mathbf{B}^{-1} \bm\Sigma\mathbf{B}
+ 2 \mathbf{B} \bm\Sigma \mathbf{B}^{-1} \bm\Sigma,
&&
\notag
\end{align}
required in the expansion of Cauchy stress,
where $\bm\Sigma = \mathbf{F}\bm\tau\mathbf{F}^\mathrm{T}$ and $\mathbf{B}=\mathbf{FF}^\mathrm{T}$.
In the reference configuration $\mathcal{B}_r$ these reduce to
\begin{eqnarray}
&& \dfrac{\partial I_1}{\partial \mathbf{F}} = 2 \mathbf{I}, \ \ \dfrac{\partial I_2}{\partial \mathbf{F}} =4\mathbf{I},\ \  \dfrac{\partial I_3}{\partial \mathbf{F}} =2\mathbf{I},\ \  \dfrac{\partial I_6}{\partial \mathbf{F}} = 2 \bm\tau, \ \  \dfrac{\partial I_7}{\partial \mathbf{F}} = 4  \bm\tau ,\ \
\dfrac{\partial I_8}{\partial \mathbf{F}} = 2 \bm\tau^2,\ \  \dfrac{\partial I_9}{\partial \mathbf{F}} = 4  \bm\tau^2.\notag
\end{eqnarray}

%+++++++++++++++++++++++++++++++++++++++

\subsection*{A2. Second derivatives of the invariants}

%+++++++++++++++++++++++++++++++++++++++

Here we present the second derivatives of the invariants in index notation, omitting the details of the
calculations. In the form required for the calculation of the components of $\bm{\mathcal{A}}_0$ we have
\begin{eqnarray}
&&
F_{p \alpha} F_{q\beta} \dfrac{\partial^2 I_1}{\partial F_{i \alpha} \partial F_{j \beta}}
 = 2 B_{pq} \delta_{ij}, \notag \\[4pt]
&&
 F_{p \alpha} F_{q\beta} \dfrac{\partial^2 I_2}{\partial F_{i \alpha} \partial F_{j \beta}}
 = 2 I_1 B_{pq} \delta_{ij} + 4B_{pi}B_{qj} - 2B_{iq}B_{jp} - 2B_{pr}B_{rq}\delta_{ij}
  - 2 B_{pq} B_{ij},\notag \\[4pt]
&&
 F_{p \alpha} F_{q\beta} \dfrac{\partial^2 I_3}{\partial F_{i \alpha} \partial F_{j \beta}}
 = 4I_3 \delta_{ip} \delta_{jq} - 2I_3 \delta_{iq} \delta_{jp}, \notag \\[4pt]
&&
 F_{p \alpha} F_{q\beta} \dfrac{\partial^2 I_6}{\partial F_{i \alpha} \partial F_{j \beta}}
 = 2 \Sigma_{pq} \delta_{ij}, \notag \\[4pt]
&&
 F_{p \alpha} F_{q\beta} \dfrac{\partial^2 I_7}{\partial F_{i \alpha} \partial F_{j \beta}}
 = 2 \Sigma_{pq} B_{ij} + 2 \Sigma_{pr}B_{rq} \delta_{ij} + 2 B_{pr} \Sigma_{rq} \delta_{ij}
  + 2 \Sigma_{ij} B_{pq} + 2 \Sigma_{pj} B_{iq} + 2 \Sigma_{qi} B_{jp}.\notag
\end{eqnarray}
The second derivatives of $I_8$ and $I_9$ can be deduced from those of $I_6$ and $I_7$,
respectively, by replacing $\bm\tau$ by $\bm\tau^{2}$, or $\bm\Sigma$ by $\bm\Sigma\mathbf{B}^{-1}\bm\Sigma$.
In the reference configuration $\mathcal{B}_r$ these reduce to
\begin{eqnarray}
&&\dfrac{\partial^2 I_1}{\partial F_{i p} \partial F_{j q}}= 2 \delta_{pq} \delta_{ij}, \quad  \dfrac{\partial^2 I_2}{\partial F_{i p} \partial F_{j q}}=2\delta_{pq}\delta_{ij}+4\delta_{ip}\delta_{jq}-2\delta_{iq}\delta_{jp},\notag\\[4pt]
&&\dfrac{\partial^2 I_3}{\partial F_{i p} \partial F_{j q}}=4\delta_{ip}\delta_{jq}-2\delta_{iq}\delta_{jp},\quad \dfrac{\partial^2 I_6}{\partial F_{i p} \partial F_{j q}}=2\tau_{pq}\delta_{ij}, \notag\\[4pt]
&&  \dfrac{\partial^2 I_7}{\partial F_{i p} \partial F_{j q}}=6\tau_{pq}\delta_{ij}+2\tau_{ij}\delta_{pq}+2\tau_{jp}\delta_{iq}+2\tau_{iq}\delta_{jp}.\notag
\end{eqnarray}

%%%%%%%%%%%%%%%%

\section*{Acknowledgements}

This work is supported by a Senior Marie Curie Fellowship awarded by the Seventh Framework Programme of the European Commission to the second author, and by an E.T.S. Walton Award given to the third author by Science Foundation Ireland.
This material is partly based upon works supported by the Science Foundation Ireland under Grant No. SFI 08/W.1/B2580.

%%%%%%%%%%%%%%%%

%%%%%%%%%%%%%%%%%%%%%%%%


\begin{thebibliography}{99}

\bibitem[Biot(1939)]{Biot39}
M.A. Biot,
Non-linear theory of elasticity and the linearized case for a body under initial stress,
Phil. Mag. \textbf{27}, 468--489 (1939).

\bibitem[Biot(1940)]{Biot40}
M.A. Biot,
The influence of initial stress on elastic waves,
J. Appl. Phys. \textbf{11}, 522--530 (1940).

\bibitem[Biot(1965)]{Biot65}
M.A. Biot,
Mechanics of Incremental Deformations,
John Wiley, New York (1965).

\bibitem[Boulanger and Hayes(1993)]{Boul93}
P. Boulanger and M. Hayes, Bivectors and Waves in Mechanics and Optics, Chapman \& Hall (1993).

\bibitem[Chadwick(1997)]{Chad97}
P. Chadwick,
The application of the Stroh formulation to prestressed elastic media,
Math. Mech. Solids \textbf{2}, 379--403 (1997).

\bibitem[Chadwick and Ogden(1971)]{Chad71}
P. Chadwick and R.W. Ogden, On the definition of elastic moduli, Arch. Ration. Mech. Anal.  \textbf{44}, 41--53 (1971).

\bibitem[Destrade et al.(2010)]{Destrade10}
M. Destrade, M.D. Gilchrist, G. Saccomandi,
Third- and fourth-order constants of incompressible soft solids
and the acousto-elastic effect,
J. Acoust. Soc. Am. \textbf{127},  2759--2763 (2010).

\bibitem[Hayes and Rivlin(1961)]{Haye61}
M. Hayes and R.S. Rivlin, Propagation of a plane wave in an isotropic elastic
material subjected to pure homogeneous deformation, Arch. Rat. Mech. Anal. \textbf{8}, 15--22 (1961).

\bibitem[Hoger(1985)]{Hoge85}
A. Hoger,
On the residual stress possible in an elastic body with material symmetry,
Arch. Rat. Mech. Anal. \textbf{88}, 271--290 (1985).

\bibitem[Hoger(1986)]{Hoge86}
A. Hoger,
On the determination of residual stress in an elastic body,
J. Elasticity \textbf{16}, 303--324 (1986).

\bibitem[Hoger(1993)]{Hoge93}
A. Hoger,
The elasticity tensors of a residually stressed material,
J. Elasticity \textbf{31}, 219--237 (1993).

\bibitem[Holzapfel and Ogden(2010)]{Holz10}
G.A. Holzapfel and R.W. Ogden,
Modelling the layer-specific 3D residual stresses in arteries, with an application to the human aorta,
J. R. Soc. Interface \textbf{7}, 787--799 (2010).

\bibitem[Hughes and Kelly(1953)]{HuKe53}
D.S. Hughes, J.L. Kelly,
Second-order elastic deformation of solids,
Phys. Rev. \textbf{92}, 1145--1149 (1953).

\bibitem[Johnson and Hoger(1993)]{John93}
B.E. Johnson,  A. Hoger,
The dependence of the elasticity tensor on residual stress,
J. Elasticity \textbf{33}, 145--165 (1993).

\bibitem[Man(1998)]{Man98}
C.-S. Man,
Hartig's law and linear elasticity with initial stress,
Inv. Prob. \textbf{14}, 313--319 (1998).

\bibitem[Man and Lu(1987)]{Man87}
C.-S. Man and W.Y. Lu,
Towards an acoustoelastic theory for measurement of residual stress,
J. Elasticity \textbf{17}, 159--182 (1987).

\bibitem[Murnaghan(1937)]{Murn}
F. D. Murnaghan, Finite deformations of an elastic solid,
Am. J. Mathematics \textbf{59},  235--260 (1937).

\bibitem[Ogden(1984)]{Ogde84}
R.W. Ogden, Non-Linear Elastic Deformations, Ellis Horwood, Chichester (1984).

\bibitem[Ogden(2003)]{Ogde03}
R.W. Ogden, Nonlinear elasticity, anisotropy and residual stresses in soft tissue, in Biomechanics of Soft Tissue in Cardiovasular Systems, edited by G.A. Holzapfel and R.W. Ogden, Springer, Wien (2003), pp.~65--108.

\bibitem[Ogden(2007)]{Ogde07}
R.W. Ogden, Incremental statics and dynamics of pre-stressed elastic materials, in Waves in Nonlinear Pre-Stressed Materials, edited by M. Destrade and G. Saccomandi, Springer, Wien (2007), pp.~1--26.

\bibitem[Ogden and Singh(2011)]{Ogde11}
R.W. Ogden and B. Singh,
Propagation of waves in an incompressible transversely isotropic elastic solid with initial stress: Biot revisited,
J. Mech. Mat. Structures, to appear (2011).

\bibitem[Saravanan(2008)]{Sara08}
U. Saravanan,
Representation for stress from a stressed reference configuration,
Int. J. Eng. Sci. \textbf{46}, 1063--1076 (2008).

\bibitem[Scott(1974)]{Scot74}
N.H. Scott, Some motions of fibre reinforced elastic materials, Ph.D. Thesis, University of East Anglia (1974).

\bibitem[Scott and Hayes(1985)]{Scot85}
N.H. Scott and M. Hayes, A note on wave propagation in internally constrained
hyperelastic materials, Wave Motion \textbf{7}, 601--605 (1985).

\bibitem[Spencer(1971)]{Spen71}
A.J.M. Spencer, Theory of invariants, in Continuum Physics Vol. 1, edited by A.C. Eringen, Academic Press, New York (1971), pp.~239--353.

\bibitem[Toupin and Bernstein(1961)]{Toup61}
R.A. Toupin and B. Bernstein, Sound waves in deformed perfectly elastic materials. Acoustoelastic effect, J. Acoust. Soc. Am. \textbf{33}, 216--225 (1961).

\bibitem[Truesdell(1961)]{True61}
C. Truesdell, General and exact theory of waves
in finite elastic strain, Arch. Rat. Mech. Anal. \textbf{8}, 263--296 (1961).

\bibitem[Zheng(1994)]{Zhen94}
Q.-S. Zheng, Theory of representations for tensor functions---a unified invariant approach to constitutive equations, Appl. Mech. Rev. \textbf{47}, 545--587 (1994).


\end{thebibliography}
\end{document}